\documentclass[aps,prc,preprintnumbers,superscriptaddress,showpacs,nofootinbib]{revtex4-1}

\newcommand{\be}{\begin{eqnarray*}}
\newcommand{\ee}{\end{eqnarray*}}
\usepackage{amsmath,graphicx}
\usepackage{amssymb}
\usepackage[english]{babel}
\usepackage{indentfirst}
\usepackage{amsxtra}
\usepackage{amsfonts}
\usepackage{bm}
\usepackage{amsmath}
\usepackage{supertabular}
\usepackage{multirow}
\usepackage[mathcal]{eucal}
\usepackage[usenames]{color}
\usepackage{bm}

\usepackage{url}
\usepackage{hyperref}%

\begin{document} 
\title{\textbf{Next-to-next-to-leading order Skyrme interaction in nuclear matter: Nuclear bulk quantities at second order in perturbation theory }}
\author{Kassem Moghrabi}
\affiliation{
Multidisciplinary Physics Lab, Lebanese University, Faculty of Sciences I, Hadath, Lebanon}
\affiliation{American University of Science and Technology, Beirut, Lebanon}
\begin{abstract}
We present the explicit form of the next-to-next-to-leading order (N$^2$LO) Skyrme interaction in momentum space by including the fourth-order gradient potentials to the standard Skyrme interaction. With the N$^2$LO Skyrme interaction, we evaluate the second-order corrections to the nuclear bulk quantities of nuclear matter: equation of state (EoS) of isospin symmetric and pure neutron matter, density-dependent in-medium effective nucleon mass, isospin-asymmetry energy, pressure and incompressibility. These second-order contributions are ultraviolet (UV) divergent due to the zero range character of the interaction and renormalized using the techniques of dimensional regularization (DR) with the minimal subtraction scheme (MS). We adjust the 18 parameters of the interaction by performing a global fit to the nuclear bulk quantities. Besides the too strong dependence $k_F^{12}$ of several second-order corrections, a very good reproduction of a realistic nuclear matter saturation curve with all the nuclear bulk quantities in the density region $0.12<\rho<0.20\;$ fm$^{-3}$ is obtained.
\end{abstract}
\pacs{21.60.Jz,21.30.-x,21.65.Mn}
\maketitle

\section{Introduction}
The energy-density-functional (EDF) theory, often referred to as self-consistent mean-field (SCMF) methods, is an effective tool for the microscopic description of medium- and heavy-mass nuclei. These methods provide the lowest-order nuclear binding energies and successfully reproduce the nuclear bulk properties for such nuclei \cite{binder}. In the EDF framework, the energy of the system is computed by using functionals of the density which are usually derived from effective interactions \cite{binder,vretenar,reinhard}. The currently used EDF's are usually produced by phenomenological interactions such as the standard Skyrme \cite{skyrme,skyrme2,vautherin} and the Gogny interactions \cite{gogny,girod}. However, these mean-field methods can not well reproduce the spectroscopic factors and the fragmentation of the single-particle states. A possible way to overcome this limitation is to go beyond the mean-field level and consider correlations to better describe nuclear observables, achieve a higher precision and increase the predictive power of such methods in exotic regions of the nuclear chart.\\
In the context of beyond mean-field methods applied to nuclear matter, second-order contributions to the equation of state (EoS) of isospin symmetric nuclear matter 
with the simplified ($t_0-t_3$) or leading-order (LO) Skyrme interaction \cite{moghrabi1,brenna} and with the standard Skyrme interaction (or the next-to-leading order (NLO) Skyrme interaction) neglecting both spin-orbit and tensor forces \cite{orsay} have been carried out. Moreover, second-order corrections to the various nuclear bulk quantities of nuclear matter with the standard Skyrme interaction including its corresponding spin-orbit and tensor forces have been pursued in \cite{kaiser1}. The adjustment of the Skyrme parameters in \cite{moghrabi1,brenna,orsay} has been performed in such a way the constraints were imposed only on the EoS of isospin symmetric nuclear matter and/or that of pure neutron matter. However, no constraints have been imposed on other bulk quantities such as the incompressibility, effective mass and isospin-asymmetry energy that are very important ingredients  to reproduce a reasonable EoS curve. Moreover, it has been concluded in \cite{kaiser1} that a good reproduction of the realistic nuclear matter saturation curve cannot be obtained upon readjusting the parameters of the standard Skyrme interaction because of the too strong density-dependence $k_F^8$ of several second-order contributions and that the adjustment procedure 
should include all possible constraints on the bulk quantities of nuclear matter.\\
The standard Skyrme interaction is a set of contact interactions with 11 parameters fitted to reproduce properties of infinite nuclear matter and some selected nuclei \cite{sly5}. 
However, this interaction is not able to reproduce the correct isovector splitting of the effective mass of Brueckner-Bethe-Goldstone results \cite{bender,baldo} and the high density behaviour of nuclear matter. Moreover, there are some nuclear properties which cannot correctly be reproduced with the standard Skyrme terms, especially as one moves away from the valley of $\beta$-stability \cite{erler}. 
Different strategies have been employed to overcome these difficulties by either adding extra density dependencies on the velocity dependent terms \cite{chamel0,xiong} or by including higher order derivative terms \cite{carlsson}. Intense work is currently being devoted to improve the existing Skyrme parametrizations, in particular considering additional terms to the standard Skyrme functionals \cite{dave1,dave2,dave3,doba1,carlsson,jeremy}. Davesne et al. \cite{dave1,dave2} have considered the Skyrme pseudo-potentials by adding the fourth- and sixth- order gradient terms to the standard Skyrme interaction. With such terms, the spin/isospin decomposition of an equation of state obtained from ab-initio methods is fairly reproduced and a more precise description of the high-density region is obtained. Moreover, the higher-order derivative terms have been shown to have a stronger influence on the high density part with a negligible modification to the low density part.\\
Our objective is to go beyond the standard Skyrme interaction by including all fourth-gradient potentials. This extended interaction, that describes the two-nucleon system at next-to-next-to-leading order (N$^2$LO) in pionless chiral effective field theory \cite{meisner}, is a set of contact forces ($s-$,$p-$, and $d-$waves) of 24 parameters consisting of zero, two and four gradient potentials as well as spin-orbit and tensor terms. The study of the N$^2$LO tensor forces (with 4 parameters) is devoted for a future work. The remaining 18 parameters of the N$^2$LO Skyrme interaction are determined by readjusting them in such a way the EoS saturation curve is well reproduced.\\
In this article, we go beyond the mean-field level and consider second order diagrams (2p-2h excitations). We derive analytically the second-order corrections to the various nuclear bulk properties: the EoS of isospin symmetric and pure neutron matter, effective mass, isospin-asymmetry energy, pressure and incompressibility using the N$^2$LO Skyrme interaction. We adjust the Skyrme parameters so that a good reproduction of the EoS curve is obtained with reasonable values of the bulk quantities at saturation. However, since the N$^2$LO Skyrme interaction is of zero range, UV divergences occur at second order in many-body perturbation theory. To remove the UV divergences, renormalization methods such as momentum cutoff (MC) or dimensional regularization (DR) with the minimal subtraction scheme (MS) are used. The UV divergent terms for the EoS of isospin symmetric matter at second order are proportional to $\Lambda^{1,3}\,k_F^{9}$ and  $\Lambda\,k_F^{11}$ if a MC scheme is used. These terms cannot be renormalized since they cannot be absorbed by a redefinition of the existing Skyrme parameters. In this work, we use DR/MS since it sets all power divergent integrals to zero, and therefore all second-order corrections to bulk quantities in nuclear matter are hence renormalized . \\
The article is organized as follows. In section 2, the standard Skyrme interaction is introduced as a zero-range contact interaction limited to second order in gradient terms ($\sim \nabla^2$). 
Next, the N$^2$LO Skyrme interaction is introduced in momentum space as a Galilei- and time-reversal-invariant two-nucleon interaction expanded in gradient terms up to fourth order ($\sim \nabla^4$). Beyond the standard Skyrme interaction, the N$^2$LO Skyrme interaction consists of fourth-gradient velocity-dependent and spin-orbit potentials that simulate finite-range effects. In refs.~\cite{dave1,dave2}, the explicit form of the N$^2$LO effective pseudo-potential is constructed given that it should be compatible with all required symmetries, and especially with gauge invariance. Based on this constraint, the spin-orbit contributions have been discarded at N$^2$LO since such contributions do not fulfil gauge invariance. However, this invariance is not explicitly required from basic principles as mentioned in \cite{dave1,dave2}, so the N$^2$LO Skyrme interaction presented in this article is not constrained to be gauge invariant. \\
In section 3, the first-order (Hartree-Fock) contributions to the nuclear bulk quantities, EoS of isospin symmetric nuclear matter and pure neutron matter, effective mass, isospin asymmetry energy and the incompressibility, are evaluated and their analytical expressions are verified to be in agreement with those derived in ref. \cite{dave3}. It is found that these first-order corrections can be written as a polynomial of odd power of the Fermi momentum $k_F$ multiplied with products of Skyrme parameters and that the spin-orbit (non-gauge invariant) terms do not contribute to the various bulk quantities at first order. \\
Section $4$ is devoted to the analytical calculations of the second-order corrections in many-body perturbation theory to the various nuclear bulk quantities. Due to the zero range character of the N$^2$LO Skyrme interaction, these many-body quantities are UV divergent at second order. The divergences appear only in the second-order diagrams in which the momentum transfer $q$ (between the incoming and scattered particles) is not constrained and can take any non-negative value. The divergent integrals are treated by using the techniques of dimensional regularization with the minimal subtraction scheme (DR/MS) and thus vanish identically. The finite parts of the second-order contributions arise entirely from integrals evaluated over three Fermi spheres. It is found that the second-order corrections to the various nuclear bulk quantities can be written as a polynomial of even powers of the Fermi momentum $k_F$ multiplied with some functions that depend on products of Skyrme parameters. Unlike the first-order contributions, it is found that the non-gauge invariant terms do contribute to the various nuclear bulk quantities at second order. Therefore, one cannot discard such terms at second order in perturbation theory.\\
Section $5$ is devoted to the adjustments of Skyrme parameters where the EoS curve provided by the phenomenological SLy5-interaction \cite{sly5} is taken as a benchmark. Single adjustments of parameters are performed so that the EoS of pure neutron matter and isospin symmetric nuclear matter are reproduced. It is found that although a very good reproduction of the EoS of isospin symmetric matter is obtained, the other bulk quantities such as the effective mass and the isospin-asymmetry energy take unreasonable values. Next, all nuclear bulk quantities are included in the fitting procedure. It is found that one can obtain a very good reproduction of the realistic nuclear matter saturation curve in the density region $0.12$ fm$^{-3}<\rho<0.20$ fm$^{-3}$. We draw conclusions in section 6. The analytical expressions for the various nuclear bulk quantities are given in the appendix.
\section{N$^2$LO Skyrme interaction}
We start by writing the transition matrix-elements of the standard Skyrme interaction in momentum-space with zero- and two-derivative contact forces \cite{skyrme,skyrme2,vautherin}:
\begin{equation} V_{\text{Sk2}} = t_0(1+x_0 P_\sigma) +{t_1 \over 2}(1+x_1 P_\sigma) 
(\vec{k}^{'2}+ \vec{k}^2)+t_2(1+x_2 P_\sigma)\, 
\vec{k}^{'}\cdot\vec{k} + i\, W_0(\vec \sigma_1+\vec \sigma_2)
\cdot (\vec{k}^{'}\times \vec{k}) .
\label{eq:vsk}
\end{equation}
In this notation, the subscript Sk2 denotes the Skyrme interaction expanded up to second order in momenta, $\vec{\sigma}_{1,2}$ are Pauli spin-matrices, $P_\sigma\!=\!(1+\vec \sigma_1\cdot\vec \sigma_2)/2$ is the spin-exchange operator, $\vec k = (\overrightarrow{\nabla}_1-\overrightarrow{\nabla}_2)/2i$ and $\vec k^{'} = (\overleftarrow{\nabla}_1-\overleftarrow{\nabla}_2)/2i$ denote the relative momentum between two nucleons; $\vec k$ ($\vec k^{'}$) acts on the wave function to its right (left). The Skyrme parameters $t_0$, $t_1$, $t_2$, $x_0$, $x_1$, $x_2$ and $W_0$ are to be determined by a fitting procedure. The standard Skyrme interaction in eq.~(\ref{eq:vsk}) consists of a central $s-$wave contact potential ($t_0$), non-local $s$-wave effective range and $p$-wave contact potentials ($t_1$) and ($t_2$) respectively, and a contact spin-orbit potential ($W_0$). There are other terms with in the standard Skyrme interaction \cite{kaiser1}: the tensor terms $(t_4)$ and $(t_5)$ that have been neglected for the sake of simplicity and the density-dependent Skyrme term $t_3\left(1+x_3P_{\sigma}\right)\rho^{\alpha}/6$ that can be included in eq.~(\ref{eq:vsk}) through the substitution of parameters:
\begin{eqnarray}
t_0\longrightarrow t_0+\frac{t_3}{6}\rho^{\alpha},\qquad t_0x_0\longrightarrow t_0x_0+\frac{t_3x_3}{6}\rho^{\alpha}.\end{eqnarray}
Such density-dependent term that comes out from a contact three body force \cite{vautherin} leads to serious problems in the calculation of many-body nuclear bulk quantities and makes conflict with the Hugenholtz-Van-Howe theorem \cite{hugen} because derivatives with respect to the density $\rho$ are involved. \\
The next-to-next-to-leading order (N$^2$LO) Skyrme interaction in momentum space is a set of two-body contact interactions with zero-, two- and four-derivative potentials. It can be written as the sum of the standard Skyrme interaction $V_{\text{Sk2}}(\vec{k},\vec{k}^{'})$ and a set of all possible fourth-derivative terms represented by $V_{\text{Sk4}}(\vec{k},\vec{k}^{'})$:
\begin{equation} 
V(\vec{k},\vec{k}^{'})=V_{\text{Sk2}}(\vec{k},\vec{k}^{'})+V_{\text{Sk4}}(\vec{k},\vec{k}^{'}),
\label{interaction}
\end{equation}
where the fourth-derivative potentials $V_{\text{Sk4}}(\vec{k},\vec{k}^{'})$ may be written as:
\begin{eqnarray} 
V_{\text{Sk4}}(\vec{k},\vec{k}^{'})&=&t_6\left(1+x_6P_{\sigma}\right)\left(\vec{k}^{'2}+\vec{k}^2\right)^2+t_7\left(1+x_7P_{\sigma}\right)\left(\vec{k}^{'}\cdot \vec{k}\right)^2+t_8\left(1+x_8P_{\sigma}\right)\left(\vec{k}^{'2}+\vec{k}^2\right)\left(\vec{k}^{'}\cdot \vec{k}\right)\nonumber\\&&\:+t_9\left(1+x_9P_{\sigma}\right)\left(\vec{k^{'}}\wedge\vec{k}\right)^2+t_{10}\left(\vec{k}^{'2}+\vec{k}^2\right)
\left(i\vec{\sigma}\cdot\vec{k}^{'}\wedge\vec{k}\right)+t_{11}\left(\vec{k}^{'}\cdot \vec{k}\right)\left(i\vec{\sigma}\cdot\vec{k}^{'}\wedge\vec{k}\right)\nonumber\\&&\: +t_{12}\left(\vec{\sigma_1}\cdot\vec{k}^{'}\wedge\vec{k}\right)
\left(\vec{\sigma_2}\cdot\vec{k}^{'}\wedge\vec{k}\right).
\label{eq:v4}
\end{eqnarray} 
The subscript Sk4 denotes the fourth-order derivative terms in the Skyrme interaction. The individual terms in eq.~(\ref{eq:v4}) stand for: $s-$wave ($t_{6}$), $p-$ wave ($t_{8},t_{10}$), $d-$ wave ($t_{11}$), and mixed $s-d$ waves ($t_{7},\,t_{9}$, $t_{12}$) contact interactions. Moreover, the first three terms ($t_6$, $t_7$ and $t_8$) represent higher-order velocity-dependent potentials as they come out from products of scalar bilinear function of momenta $F\left[\left(\vec{k}^{'2}+\vec{k}^2\right),\;\left(\vec{k}^{'}\cdot \vec{k}\right)\right]$. Whereas the last four terms ($t_9$, $t_{10}$, $t_{11}$ and $t_{12}$) are higher-order spin-orbit interactions \cite{dave1,ring}. The potentials proportional to $t_9$ and $t_{12}$ arise from products of scalar bilinear function of momenta  $G\left(i\vec{\sigma}_1\cdot\vec{k}^{'}\wedge\vec{k}\;,{i}\vec{\sigma}_2\cdot\vec{k}^{'}\wedge\vec{k}\right)$ using the well known formula for the Pauli matrices $\left(\vec{\sigma}\cdot \vec{A}\right)\left(\vec{\sigma}\cdot \vec{B}\right)=\vec{A}\cdot \vec{B}+i\vec{\sigma}\cdot \left(\vec{A}\wedge \vec{B}\right).$ The other potentials $t_{10}$ and $t_{11}$ arise from products of the LO spin-orbit potential (from the standard Skyrme interaction) with the scalar bilinear function $H$:  $\left(i\sigma\cdot\vec{k}^{'}\wedge\vec{k}\right) H\left[\left(\vec{k}^{'2}+\vec{k}^2\right),\left(\vec{k}^{'}\cdot \vec{k}\right)\right]$.\\
Three important remarks are to be considered concerning the form of the N$^2$LO Skyrme interaction. In eq.~(\ref{eq:v4}), the bilinear expressions $\left(\vec{k}^{'2}+\vec{k}^2\right)^2$ and $\left(\vec{k}^{'}\cdot \vec{k}\right)^2$ appear with different weights proportional to the Skyrme parameters $t_6$ and $t_7$.  However, these bilinear expressions appear with  weights $1:4$ in \cite{dave1} because the parameters $t_6$ and $t_7$ were chosen to be linearly dependent: $t_7=4t_6$ and $x_7=x_6$. This linear dependence arises from the fact that the authors have constructed the fourth-derivative contact interactions in  such a way their contributions to the EoS and Landau parameters of isospin symmetric nuclear matter take the same form as those derived from the standard Skyrme interaction. Moreover, the higher-order spin-orbit contributions are absent in the pseudo Skyrme interaction in \cite{dave1} because such terms are not gauge invariant and do not contribute to the first-order corrections to the various nuclear bulk quantities. However, gauge invariance is not explicitly required from basic principles as mentioned in \cite{dave1} and such non-gauge invariant terms do contribute at the second-order in perturbation theory \cite{kaiser3} and as we shall see in section 4. Therefore, for a general treatment of contact interactions at second-order, it is necessary to consider all fourth-derivative terms as in eq.~(\ref{eq:v4}). Furthermore, Skyrme \cite{skyrme2} suggested the addition of a $d-$wave term \footnote{It is pointed in \cite{dave1} that this term is not a pure $d-$ wave but also contains an $s-$wave contribution.} to the standard Skyrme interaction of the form: $\left[\vec{k}^{'2}\vec{k}^{2}-\left(\vec{k}^{'}\cdot \vec{k}\right)^2\right]$. However, by inspecting eq.~(\ref{eq:v4}), it is easy to see that the terms $\left(\vec{k}^{'2}\vec{k}^{2}\right)$ and $\left(\vec{k}^{'}\cdot \vec{k}\right)^2$ appear in the bilinear expressions proportional to $t_6$ and $t_7$ with different weights in contrary to the equal and opposite weights proposed in \cite{skyrme2}. The inclusion of N$^2$LO tensor forces and density-dependent terms are devoted for a future work \cite{moghrabi}.\\
\section{Nuclear bulk quantities at first order}
In many-body perturbation theory, the first-order contributions to the EoS of isospin symmetric nuclear matter can be written as \cite{kaiser3}:
\begin{equation} 
\Delta E^{(1)}(k_F)=\frac{1}{2\rho}\frac{1}{(2\pi)^6}\sum_{\sigma,\tau}\int d^3\vec{k}_1d^3\vec{k}_2\;\big<\vec{k}_1,\vec{k}_2\big|\bar{V}\big|\vec{k}_1,\vec{k}_2\big>\Theta(k_F-|\vec{k}_{1}|)\Theta(k_F-|\vec{k}_{2}|),
\end{equation} 
where $\bar V=\left(1-P_{\sigma}P_{\tau}P_{M}\right)V$ denotes the antisymmetrized interaction, $k_F$ is the Fermi momentum related to the total nucleon density via $k_F^3=3\pi^2\rho/2$ and the product of step functions account for the presence of the filled Fermi sea. After  summing over the spin and isospin states and integrating over two Fermi spheres, the EoS of isospin symmetric nuclear matter at first order $E^{(1)}$reads:
\begin{equation} E^{(1)}(k_F) = \frac{3\hbar^2 k_F^2}{10m}+{k_F^3 \over 4\pi^2} \bigg\{ t_0 
+{k_F^2 \over 10}\big[3t_1+t_2(5+4x_2)\big]+\frac{3k_F^{4}}{70}\left[12t_6+3t_7+2t_8\left(5+4x_8\right)\right]\bigg\} \,.
\label{e1}
\end{equation}
An important observation is that the terms proportional to $t_6$, $t_7$ and $t_8$ are reproduced as done in \cite{dave1}. Moreover, the non-gauge invariant potentials do not contribute at first-order since the cross product $\left(\vec{k}^{'}\wedge\vec{k}\right)$ vanishes at first order. Furthermore, eq.~(\ref{e1}) involves the same parameter combinations for $t_8(x_8)$ and $t_2(x_2)$ and if $t_7=4t_6$ and $x_7=x_6$ as done in \cite{dave1}, the parameter combinations for $t_1(x_1)$ and $t_6(x_6)$ become similar. \\
In the same way, the EoS of pure neutron matter at first order comes out as:
\begin{equation} E^{(1)}(k_n) =\frac{3\hbar^2 k_n^2}{10m}+ {k_n^3 \over 4\pi^2} \bigg\{ 
{t_0 \over 3}(1-x_0)+{k_n^2 \over 10}\big[t_1(1-x_1)+3t_2(1+x_2)\big] +\frac{3k_n^{4}}{70}\left[4t_6\left(1-x_6\right)+t_7\left(1-x_7\right)+6t_8\left(1+x_8\right)\right] \bigg\},
\label{en1}
\end{equation}
with $k_n$ the neutron Fermi momentum related to the neutron density by 
$\rho_n= k_n^3/3\pi^2$.\\
The density-dependent isospin asymmetry $A(k_F)$ of nuclear matter is defined as the second-derivative of the EoS of isospin-asymmetric energy evaluated at $\delta=0$:
\begin{equation} E_{\text{as}}(k_p,k_n)=  E(k_F)+\delta^2 
A(k_F)+{\cal O}(\delta^4)  \,,\end{equation}
where the isospin-asymmetry parameter $\delta$ is defined as: $ k_{n,p}^3=(1\pm \delta)k_F^3.$ $\delta=0$ corresponds to isospin symmetric nuclear matter, while $\delta=1$ corresponds to pure neutron matter.
The first-order contribution to $A(k_F)$ is given by:
\begin{equation} A^{(1)}(k_F) =\frac{\hbar^2 k_F^2}{6m}+ {k_F^3 \over 12\pi^2} \bigg\{ -t_0
(1+2x_0)+{k_F^2\over 3}\big[t_2(4+5x_2)-3t_1 x_1\big]+\frac{k_F^{4}}{6}\left[4t_6\left(1-4x_6\right)+t_7\left(1-4x_7\right)+2t_8\left(7+8x_8\right)\right]\bigg\}\,.
\label{a1}
\end{equation}
Another important property for nuclear matter is the density-dependent in-medium effective nucleon mass $M^*(k_F)$.  By adding a test-particle with momentum $\vec p$ to the filled Fermi sea, via the substitution \cite{kaiser3,pionpot}: 
\begin{eqnarray} 
\Theta(k_F-|\vec p_j|)\,\,\to\,\,
\Theta(k_F-|\vec p_j|)+ 4\pi^3 \eta \,\delta^3(\vec p_j-\vec p\,)\,,
\label{eta}
\end{eqnarray} 
with $\eta$ an infinitesimal parameter, the interacting energy per particle $E(k_F)$ changes in linear order of $\eta$ as:
\begin{eqnarray}
{k_F^3 \over 3\pi^2} E(k_F) \,\,\to \,\, {k_F^3 \over 3\pi^2}
E(k_F) + \eta U(p,k_F)+\mathcal{O}\left(\eta^2\right) \,, 
\label{singleenergy}
\end{eqnarray} 
where $U(p,k_F)$ denotes the complex-valued single-particle potential. The slope of the real part of the single-particle potential $\Re\left[U(p,k_F)\right]$ at the Fermi surface $p=k_F$ determines the density-dependent effective mass $M^*(k_F)$ of the quasi-particle excitations at the Fermi surface. The corresponding relation for 
the effective mass $M^*(k_F)$ reads:
\begin{equation}  \frac{M}{M^*(k_F)} =1+R(k_F),\qquad R(k_F)=\frac{M}{k_F} \,{\partial \Re\left[U(p,k_F)\right] \over \partial p}\bigg|_{p=k_F},
 \end{equation} 
 and $M$ the free nucleon mass. The first-order contribution to $R(k_F)$ reads:
\begin{equation}R^{(1)}(k_F) = {M k_F^3 \over 12\pi^2} \bigg\{3t_1+t_2(5+4x_2)+k_F^{2}\left[12t_6+3t_7+2t_8\left(5+4x_8\right)\right]\bigg\}\,.
\label{m1}
\end{equation}
It is clear from eqs.(\ref{e1}, \ref{en1}, \ref{a1}, \ref{m1}) that the powers of $k_F$ are odd because the Skyrme interaction comes out in even power of $k$ or $k^{'}$ and first-order diagrams (divided by the total density $\rho$) contribute by $k_F^3$. Moreover, by taking $t_7=4t_6,\;x_7=x_6$ and $t_i=x_i=0$ (for $i=9,\,10,\,11$ and $12$), it is easy to reproduce the first-order contributions derived in \cite{dave1}. Furthermore, the non-gauge invariant terms $t_9$, $t_{10}$, $t_{11}$ and $t_{12}$ do not contribute to the various nuclear bulk quantities at first order. However, they do contribute at second order as we shall see in the next section. 

\section{Nuclear bulk quantities at second order}
In this section, we derive the analytical expressions $\Delta F^{(2)}$ for the second-order contributions to the nuclear bulk quantity $F$ from the N$^2$LO Skyrme interaction defined in eq.~(\ref{interaction}). Since second-order corrections are proportional to the interaction squared, we write: $V^2=\left|V_{\text{Sk2}}+V_{\text{Sk4}}\right|^2= |V_{\text{Sk2}}|^2+\left(|V_{\text{Sk4}}|^2+V^{*}_{\text{Sk2}}V_{\text{Sk4}}+V^{*}_{\text{Sk4}}V_{\text{Sk2}}\right)$. The first term $|V_{\text{Sk2}}|^2$ corresponds to the second-order contributions $\Delta F^{(2)}_{\text{Sk2}}$ derived from the standard Skyrme interaction and their analytical expressions have been computed in ref.~\cite{kaiser1} and shown in the appendix. The other term $\left(|V_{\text{Sk4}}|^2+V^{*}_{\text{Sk2}}V_{\text{Sk4}}+V^{*}_{\text{Sk4}}V_{\text{Sk2}}\right)$ corresponds to the second-order contributions $\Delta F^{(2)}_{\text{Sk4}}$ derived from the interaction $V_{\text{Sk4}}$ and from the interference terms between $V_{\text{Sk2}}$ and $V_{\text{Sk4}}$. In terms of equations, the second-order contributions $\Delta F^{(2)}$ to the nuclear bulk quantity $F$ derived from the N$^2$LO Skyrme interaction can be written as: 
\begin{eqnarray}
\Delta F^{(2)}=\Delta F^{(2)}_{\text{Sk2}}+\Delta F^{(2)}_{\text{Sk4}}.
\end{eqnarray}
Therefore, the total contribution to a given bulk quantity $F$ at second order can be written as:
\begin{equation}
F^{(2)}=F^{(1)}+\Delta F^{(2)}.
\end{equation}
We start with the EoS of isospin symmetric nuclear matter. At second order in perturbation theory, its expression is given by:
\begin{eqnarray} 
\Delta E^{(2)}&=&\frac{1}{4\rho}\frac{1}{(2\pi)^9}\sum_{\sigma,\tau}\int d^3\vec{k_1}d^3\vec{k_2}d^3\vec{q}\;\frac{\big|\big<\vec{k_1},\vec{k_2}\big|\bar{V}\big|\vec{k_1}+\vec{q},\vec{k_2}-\vec{q}\big>\big|^2}{\epsilon_{k_1}+\epsilon_{k_2}-\epsilon_{k_1+q}+\epsilon_{k_2-q}}\Theta(k_F-|\vec{k}_1|)\Theta(k_F-|\vec{k}_2|)\nonumber\\&&\:\times\Theta(|\vec{k}_{1}+ \vec{q}|-k_F)\Theta(|\vec{k}_{2}-\vec{q}|-k_F).
\label{energysecond}
\end{eqnarray}
In eq.~(\ref{energysecond}), the terms in the denominator represent the kinetic energies of the nucleons: $\epsilon_k=\frac{\hbar^2 k^2}{2M}$ where $M$ is the free nucleonic mass. 
The first two step functions $\Theta(k_F-|k_{1,2}|)$ constrain $k_1$ and $k_2$ to lie inside the Fermi sphere and the other two step functions $\Theta(|\vec{k}_{1,2}\pm\vec{q}|-k_F)$ 
can be decomposed into three parts as done in \cite{kaiser3}:
\begin{eqnarray}
\Theta(|k_{1}+ q|-k_F)\Theta(|k_{2}- q|-k_F)=1-\left[\Theta(k_F-|k_{1}+ q|)+\Theta(k_F-|k_{2}- q|)\right]+\Theta(k_F-|k_{1,2}\pm q|).
\label{theta}
\end{eqnarray}
It is worth to mention that the first term in eq.~(\ref{theta}) corresponds to the vacuum contribution (or two-medium insertions) to the EoS because it gives no constraints on the relative momentum $q$. 
In this case, the corresponding second-order contributions are UV divergent due to the zero range character of the interaction in eq.~(\ref{interaction}). After performing the angular integrations, one gets power divergent integrals of the form $\int_0^\infty\!dq\{1,q^2,q^4,q^6,q^8\}$. In the cutoff regularization, these divergences are proportional to odd powers of the cutoff 
$\Lambda$ and the Fermi momentum $k_F$: \begin{eqnarray}
\left(\Lambda,\,\Lambda^3,\,\Lambda^5,\,\Lambda^7,\,\Lambda^9\right) k_F^{3},\;\left(\Lambda,\,\Lambda^3,\,\Lambda^5,\,\Lambda^7\right) k_F^{5}\;,\left(\Lambda,\,\Lambda^3,\,\Lambda^5\right) k_F^{7},\; \left(\Lambda,\,\Lambda^3\right) k_F^{9}\;,\Lambda\; k_F^{11}.
\label{divergences}
\end{eqnarray}
In eq.~(\ref{divergences}), the first two $k_F$-dependence could be absorbed into the coupling constants $t_{0}$, $t_1$ and $t_2$ while the third one could be absorbed into the coupling constants $t_{6}$, $t_7$ and $t_8$. It is, however, not possible to absorb the last two divergent terms into the existing coupling constants because there are no mean-field contributions proportional to $k_F^{9}$ and $k_F^{11}$. Thus to renormalize such terms, it is required to introduce a sixth-derivative two-body contact-interaction (or a four-body contact interaction) and an eighth-derivative two-body contact interaction respectively. However, the inclusion of such terms are problematic since their second-order contributions to the EoS of isospin symmetric nuclear matter give divergent terms that could be renormalized by adding higher-order derivative terms and thus making the renormalization procedure uncontrolled. This issue  reflects a problem of perturbativity of the Skyrme interaction in nuclear matter \footnote{Work is in progress including the effects of pions and overcome the non-perturbativity problem \cite{moghrabi}.}. However, using the techniques of dimensional regularization (DR) with the minimal subtraction scheme (MS), all power divergent terms are automatically set to zero and consequently all terms in eq.~(\ref{divergences}) vanish. In eq.~(\ref{theta}), the second and third terms give rise to medium effects and are called the three-medium and four-medium insertions respectively. Note that the four-medium insertion contribution to the EoS vanishes identically after performing suitable change of variables. Therefore the only finite contributions to the EoS of isospin symmetric nuclear matter comes from the three-medium insertions.\\ 
After summing over the spin and isospin states and evaluating the integrals over three Fermi spheres, it is straight forward to write the second-order contributions to the EoS of isospin symmetric nuclear matter $\Delta  E_{\text{Sk4}}^{(2)}(k_F)$ using DR/MS as: 
\begin{eqnarray}
\Delta E_{\text{Sk4}}^{(2)}(k_F) =\frac{M}{\hbar^2\pi^4}\left(e_{1s}k_F^{12}+e_{2s}k_F^{10}+e_{3s}k_F^{8}\right).
\label{eq:second}
\end{eqnarray}
In eq.~(\ref{eq:second}), the quantities $e_{is}$ correspond to the finite part of the EoS of isospin symmetric nuclear matter. $e_{1s}$ represent all the second-order contributions due to $V_{\text{Sk4}}^2$ only, whereas $e_{2s}$ represent the contributions that arise from the product of $V_{\text{Sk4}}$ with the $t_1$, $t_2$ and $W_0$ terms of the standard Skyrme interaction. Finally, $e_{3s}$ represent the contributions that arise from products of $V_{\text{Sk4}}$ with the $t_0$ term of $V_{\text{Sk2}}$. The analytical expressions of $e_{is}$ are found in the Appendix. It is to be noted that $e_{is}$ depend on the products of Skyrme parameters and numerical coefficients of the form $\ln2+a_i$, where $a_i$ is some rational number. Moreover, the potentials in eq.~(\ref{interaction}) proportional to $t_9$, $t_{10}$, $t_{11}$ and $t_{12}$ do contribute at the second-order. As a side remark in the expressions of $e_{is}$, the $s-$(or $d-$) and $p-$wave contact-interactions do not interfere in the EoS of isospin symmetric nuclear matter as shown in \cite{kaiser1}. \\
The analytical expressions for the pressure $P(k_F)$ and the incompressibility $K(k_F)$ can be derived from the expression of the EoS of isospin symmetric nuclear matter via:
\begin{eqnarray}
P=\rho^2\frac{d E}{d\rho}=\frac{2k_F^4}{9\pi^2}\frac{d E}{d k_F},\quad K = 9\frac{d P}{d\rho} = k_F^2\, {d^2 E \over d k_F^2}+4k_F\, {d E\over d k_F}.
\end{eqnarray}
In the same way, but leaving out the isospin trace, one arrives at the EoS of pure neutron matter:
\begin{eqnarray}
\Delta E_{\text{Sk4}}^{(2)}(k_n)=\frac{M}{\hbar^2\pi^4}\left(e_{1n}k_n^{12}+e_{2n}k_n^{10}+e_{3n}k_n^{8}\right),
\label{en2}
\end{eqnarray}
where $e_{in}$ are the finite parts of the EoS of pure neutron matter. The second-order contribution to the isospin asymmetry energy $A(k_F)$ is calculated by evaluating the second derivatives with respect to isospin parameter $\delta$ at $\delta =0$ of the EoS for isospin asymmetric nuclear matter. Its complete expression is given by: 
\begin{eqnarray}
\Delta A^{(2)}_{\text{Sk4}}(k_F)=\frac{M}{\hbar^2\pi^4}\left(a_1k_F^{12}+a_2k_F^{10}+a_3k_F^{8}\right),
\label{a2}
\end{eqnarray}
where $a_{i}$ are the finite parts of $A(k_F)^{(2)}$. To calculate the second-order contributions of the effective mass $M^{*}(k_F)$, the second-order interacting energy per particle in eq.~(\ref{energysecond}) is expanded to linear order of $\eta$. The complex-valued single-particle potential at second order becomes proportional to the product of three steps functions and the external momenta $\vec k_1$ and $\vec k_2$ are not constrained to lie only inside a Fermi sphere of radius $k_F$. The second-order contribution to the effective mass is given by:
\begin{eqnarray}
\Delta R^{(2)}_{\text{Sk4}}(k_F)=\frac{M^2}{\hbar^4\pi^4}\left(r_1k_F^{10}+r_2k_F^{8}+r_3k_F^{6}\right),
\label{m2}
\end{eqnarray}
where the expressions for $r_{i}$ are found in the appendix.\\
It is clear from eqs.(\ref{eq:second}, \ref{en2}, \ref{a2}, \ref{m2}) that the powers of $k_F$ are even due to the fact that the Skyrme interaction comes out in even powers of $\vec{k}$ or $\vec{k^{'}}$ and that the second-order energy per particle contributes by $k_F^4$. Unlike the first-order contributions, all Skyrme parameters contribute to the various nuclear bulk quantities at second order. 

\section{Provisional fits}
\begin{figure}[h!]
\begin{center}
\includegraphics[scale=0.45 ]{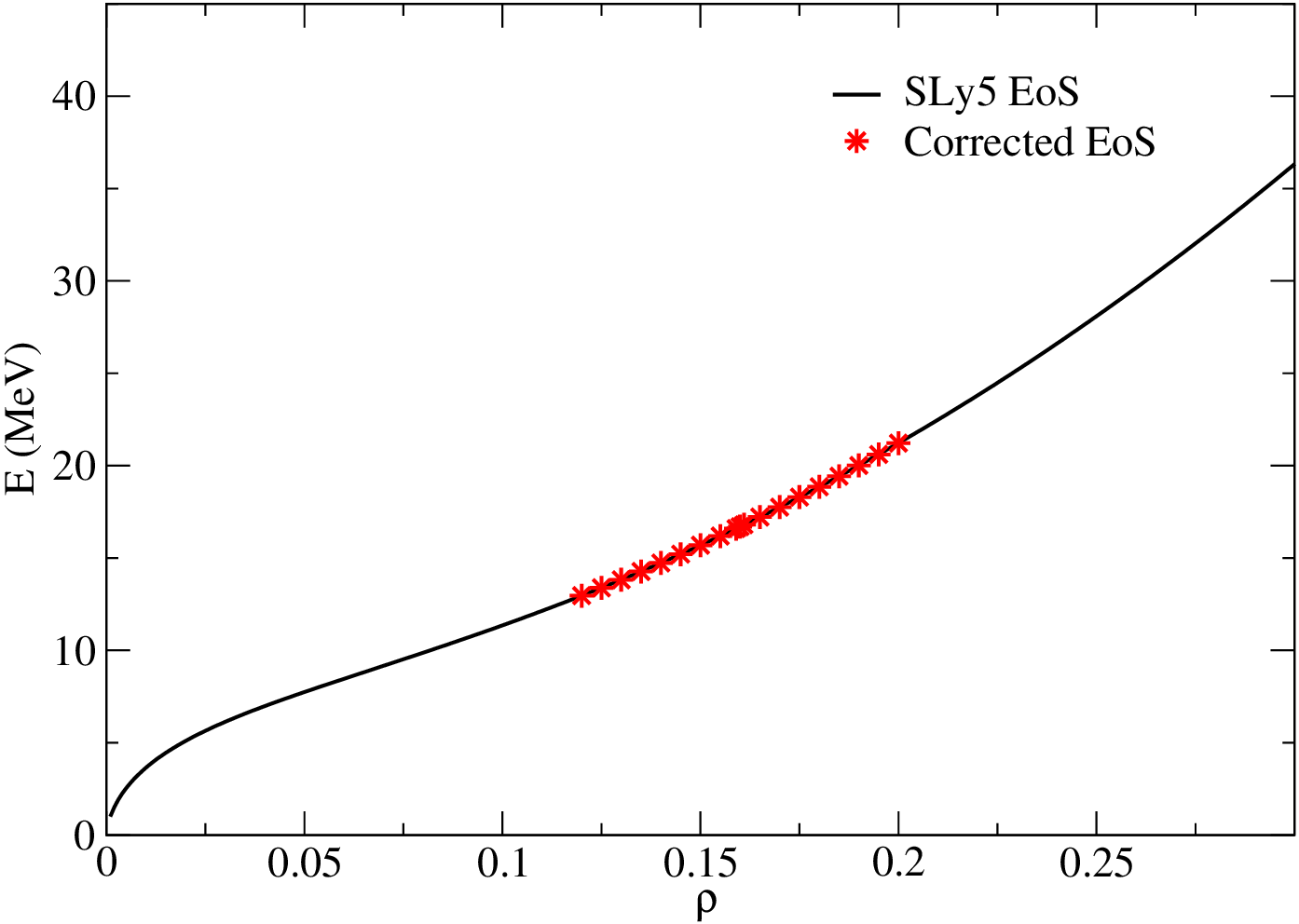}
\end{center}
\vspace{-.6cm}
\caption{\label{fig:purealone} (Color online) Fit1: Second-order refitted EoS compared with the SLy5-mean-field EoS for pure neutron matter.}
\end{figure}

\begin{figure}[h!]
\begin{center}
\includegraphics[scale=0.45 ]{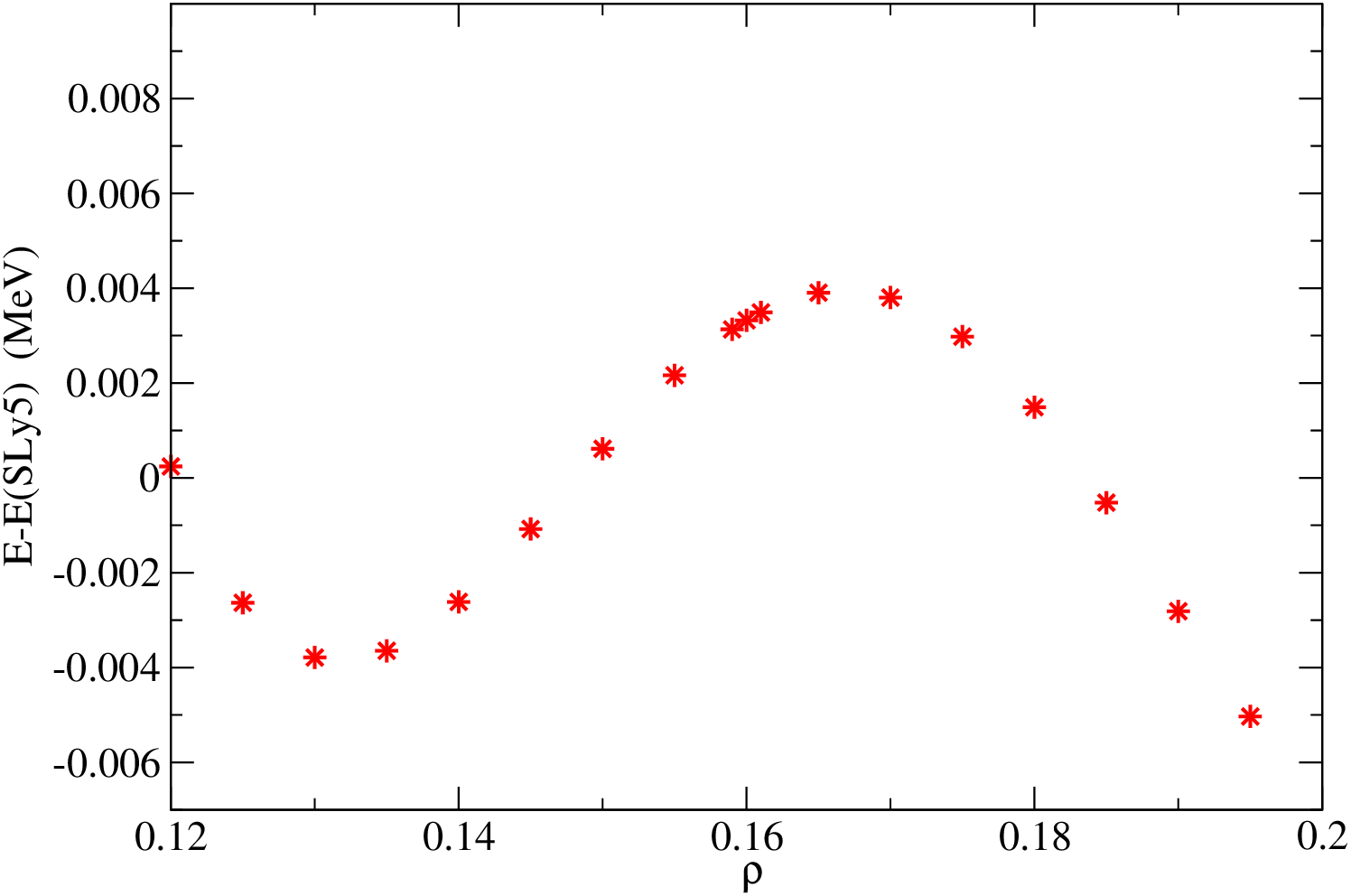}
\end{center}
\vspace{-.6cm}
\caption{\label{fig:fit-pure-alone-diff} (Color online) Fit1: Difference between the refitted second-order and the mean-field EoS for pure neutron matter.}
\end{figure}

\begin{figure}[h!]
\begin{center}
\includegraphics[scale=0.45 ]{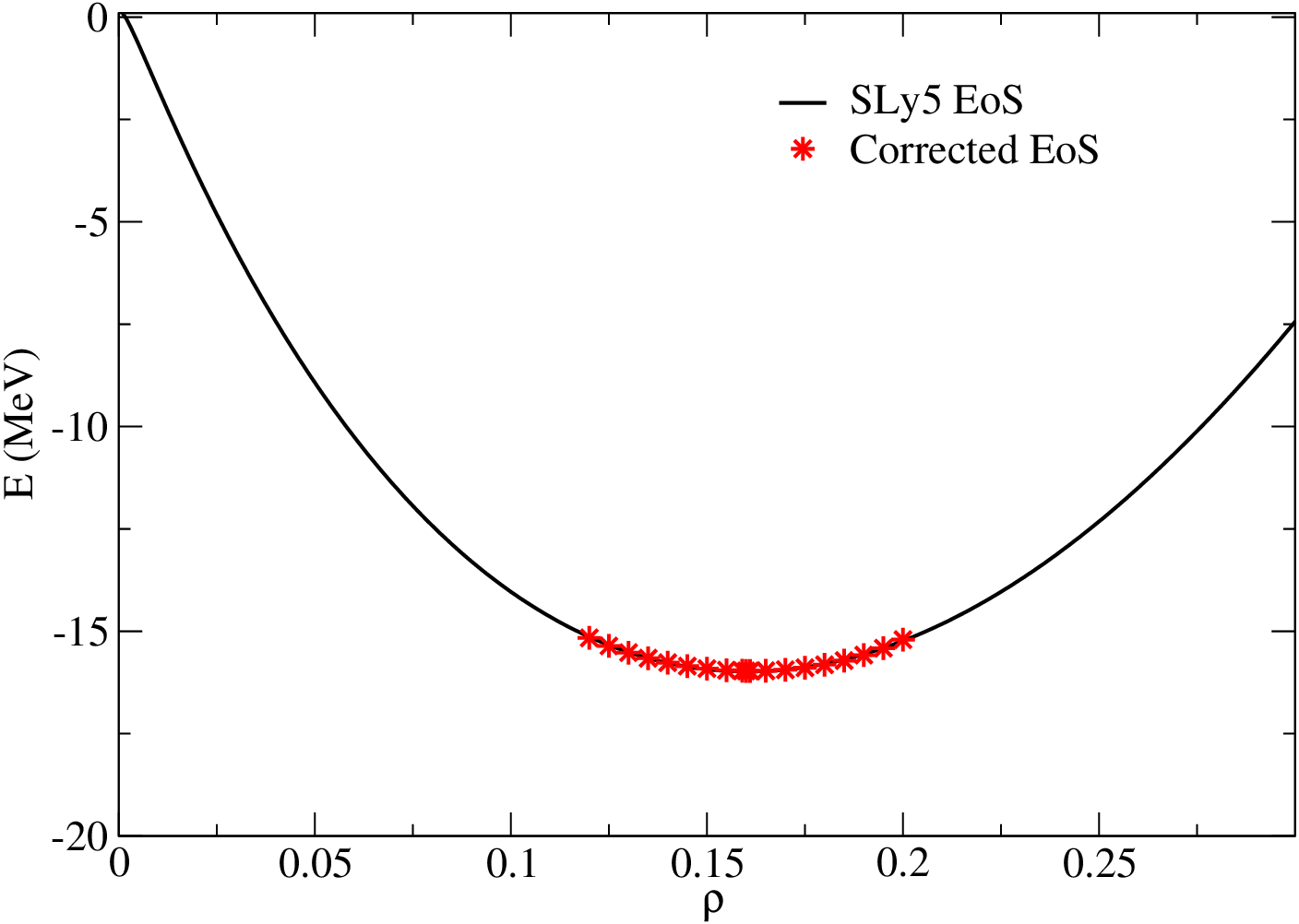}
\end{center}
\vspace{-.6cm}
\caption{\label{fig:symalone} (Color online) Fit2: Second-order refitted EoS compared with the SLy5-mean-field EoS for isospin symmetric nuclear matter.}
\end{figure}

\begin{figure}[h!]
\begin{center}
\includegraphics[scale=0.45 ]{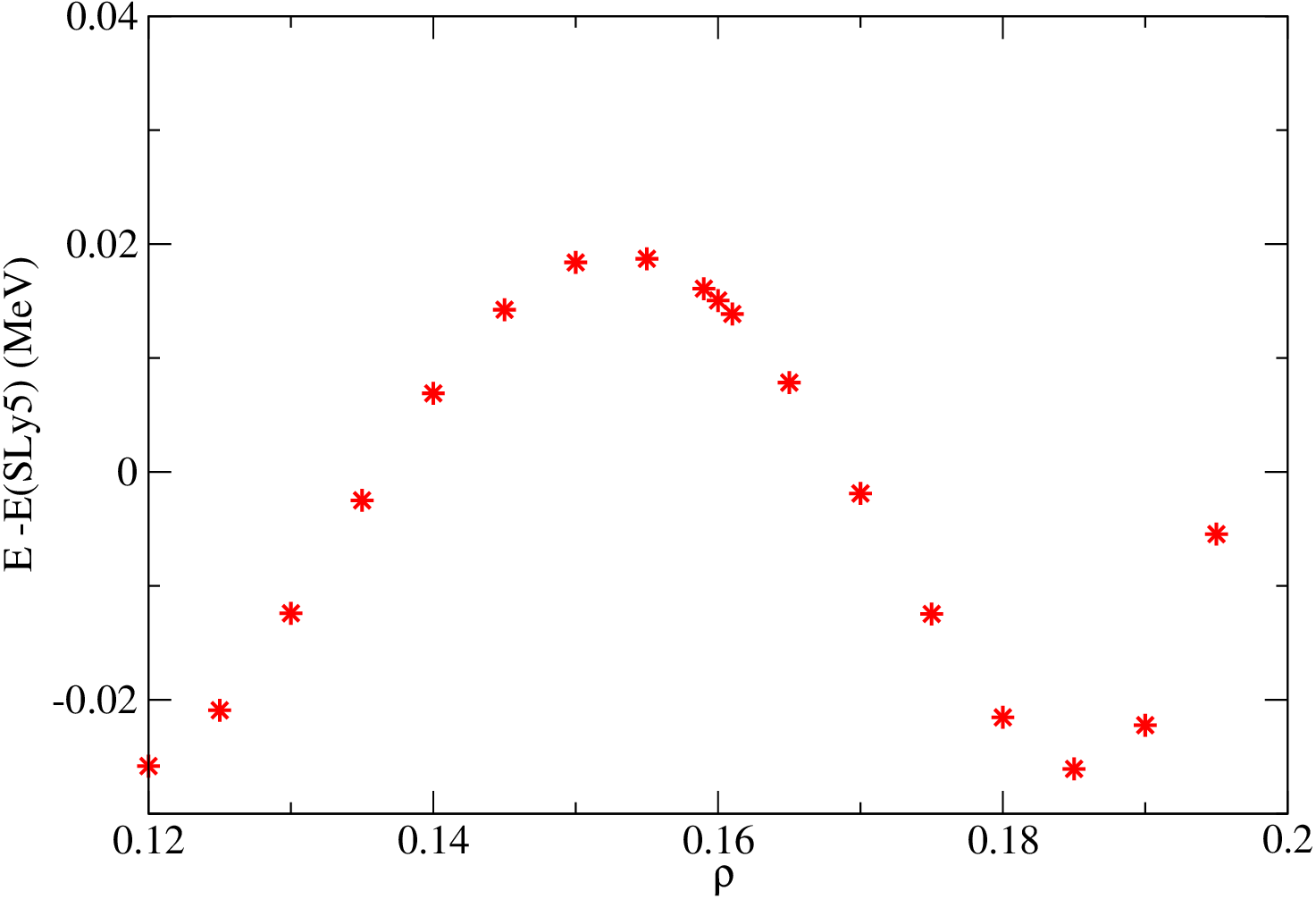}
\end{center}
\vspace{-.6cm}
\caption{\label{fig:fit-symmetric-alone-diff} (Color online) Fit2: Difference between the refitted second-order and the mean-field EoS for isospin symmetric nuclear matter.}
\end{figure}

\begin{figure}[h!]
\begin{center}
\includegraphics[scale=0.45 ]{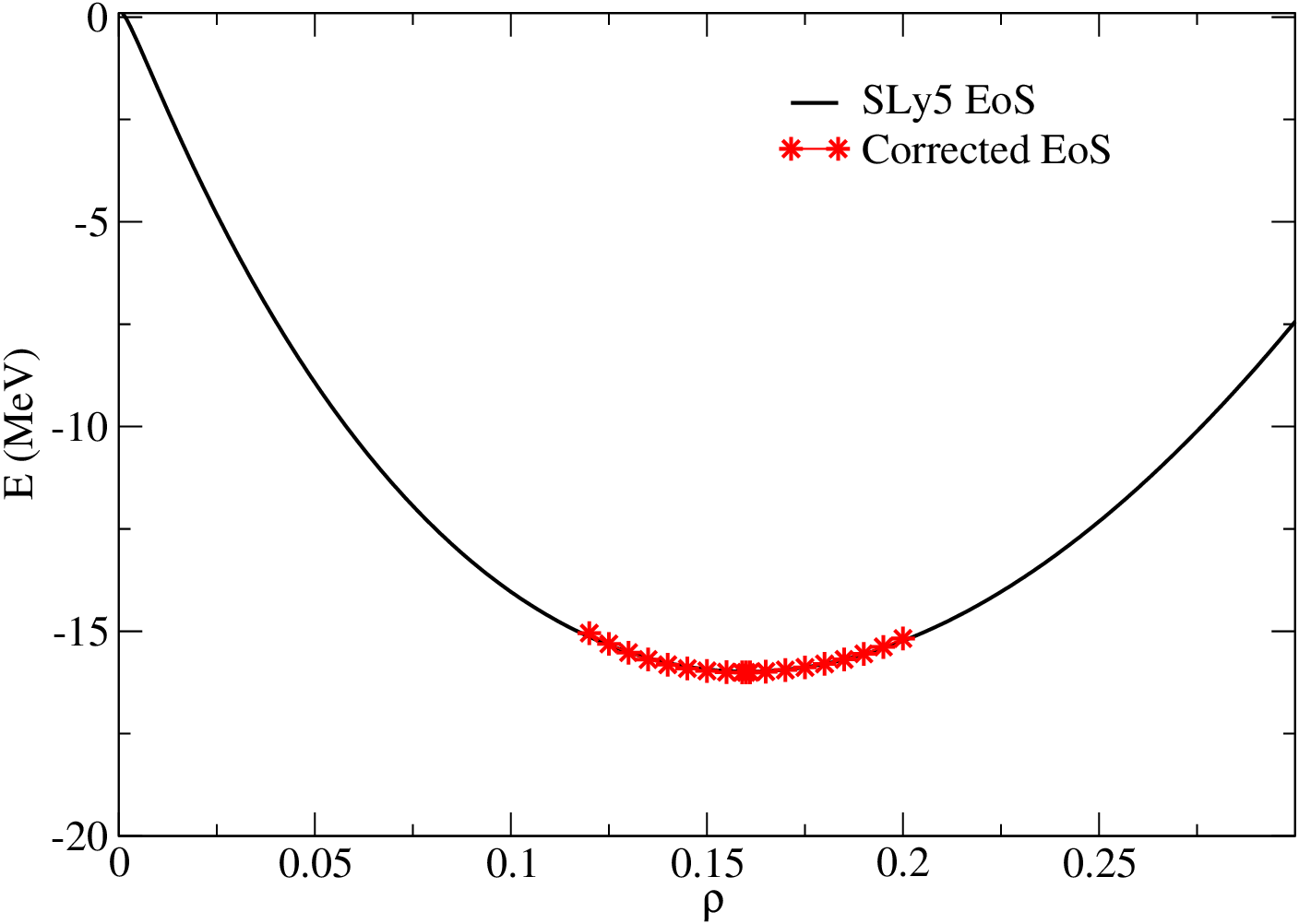}
\end{center}
\vspace{-.6cm}
\caption{\label{fig:fit-symmetric-pure1} (Color online) Fit3: Second-order refitted EoS compared with the SLy5-mean-field EoS for isospin symmetric nuclear matter.}
\end{figure}

\begin{figure}[h!]
\begin{center}
\includegraphics[scale=0.45 ]{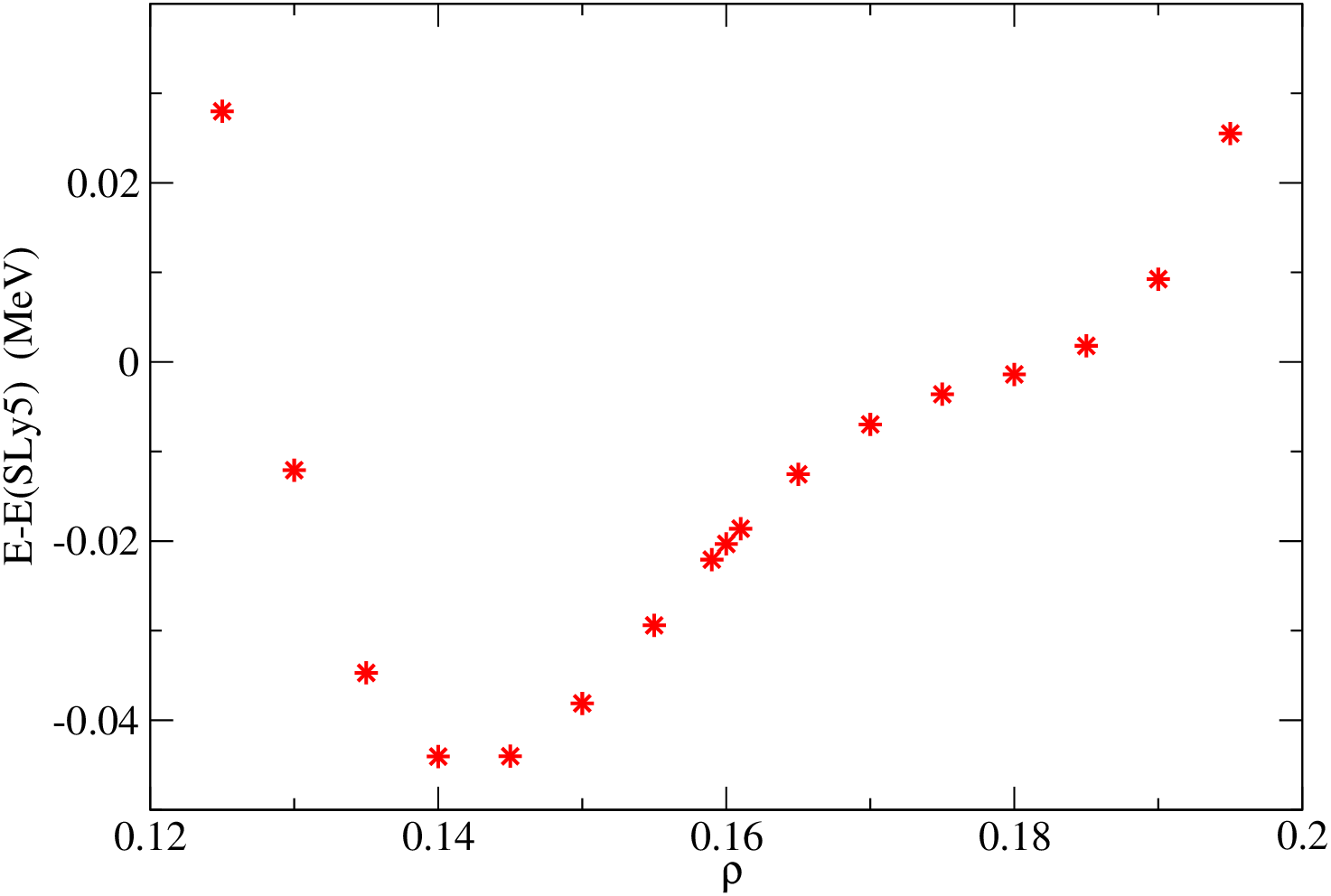}
\end{center}
\vspace{-.6cm}
\caption{\label{fig:fit-symm-pure-diff1} (Color online) Fit3: Difference between the refitted second-order and the mean-field EoS for isospin symmetric nuclear matter.}
\end{figure}

\begin{figure}[h!]
\begin{center}
\includegraphics[scale=0.45 ]{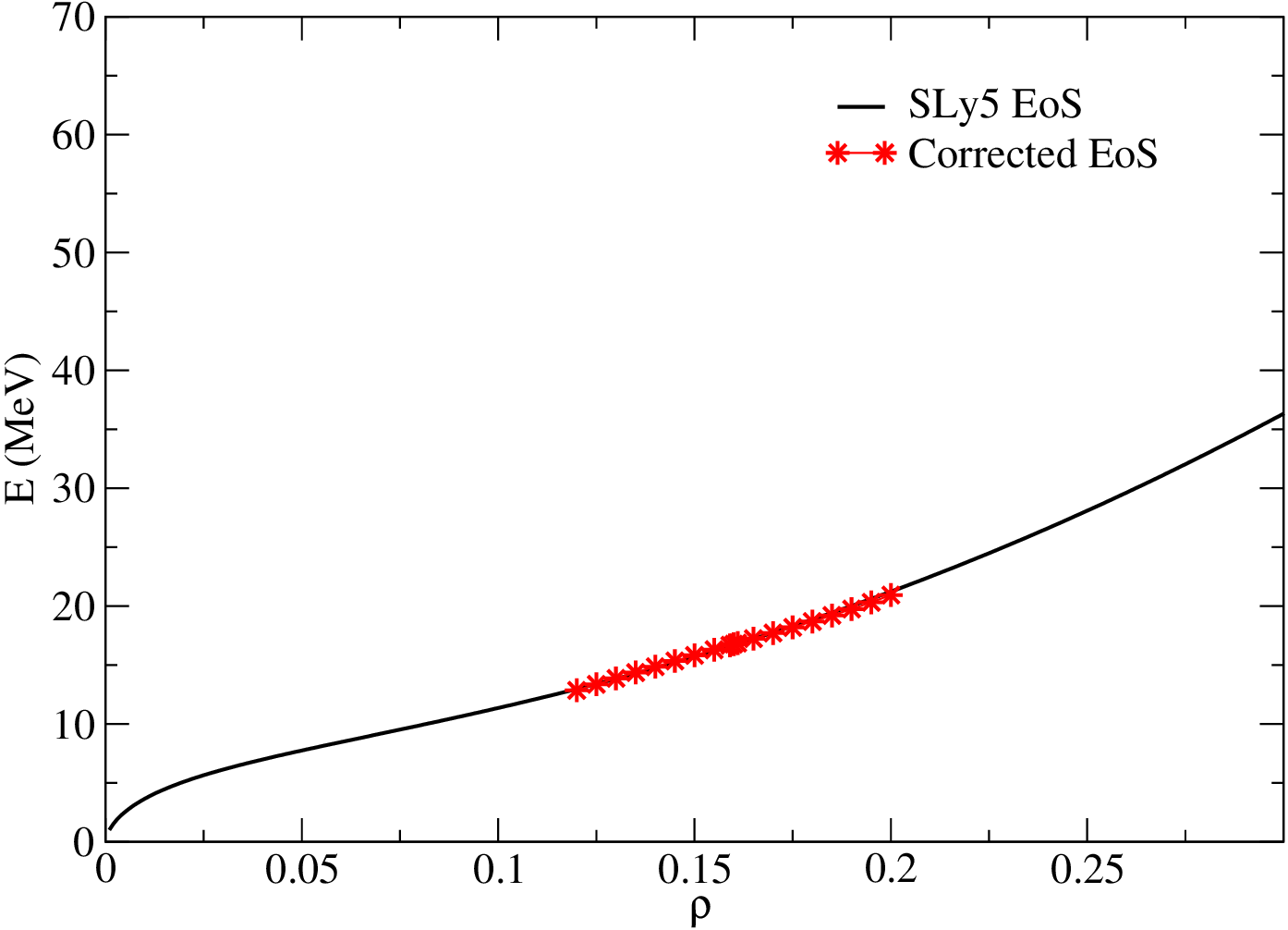}
\end{center}
\vspace{-.6cm}
\caption{\label{fig:fit-symm-pure2} (Color online) Fit3: Second-order refitted EoS compared with the SLy5-mean-field EoS for pure neutron matter.}
\end{figure}

\begin{figure}[h!]
\begin{center}
\includegraphics[scale=0.45 ]{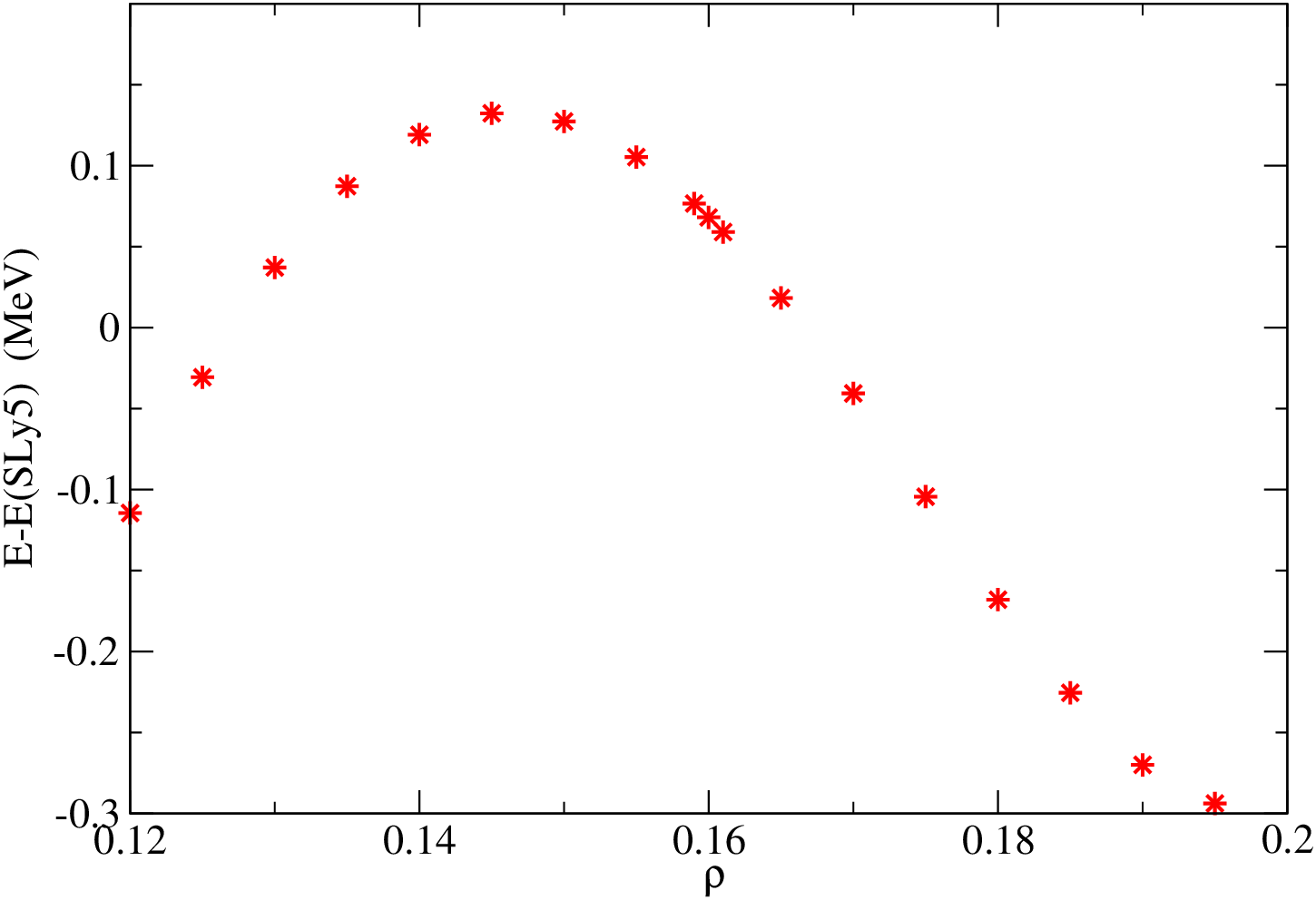}
\end{center}
\vspace{-.6cm}
\caption{\label{fig:fit-symm-pure-diff2} (Color online) Fit3: Difference between the refitted second-order and the mean-field EoS for pure neutron matter.}
\end{figure}

\begin{figure}[h!]
\begin{center}
\includegraphics[scale=0.45 ]{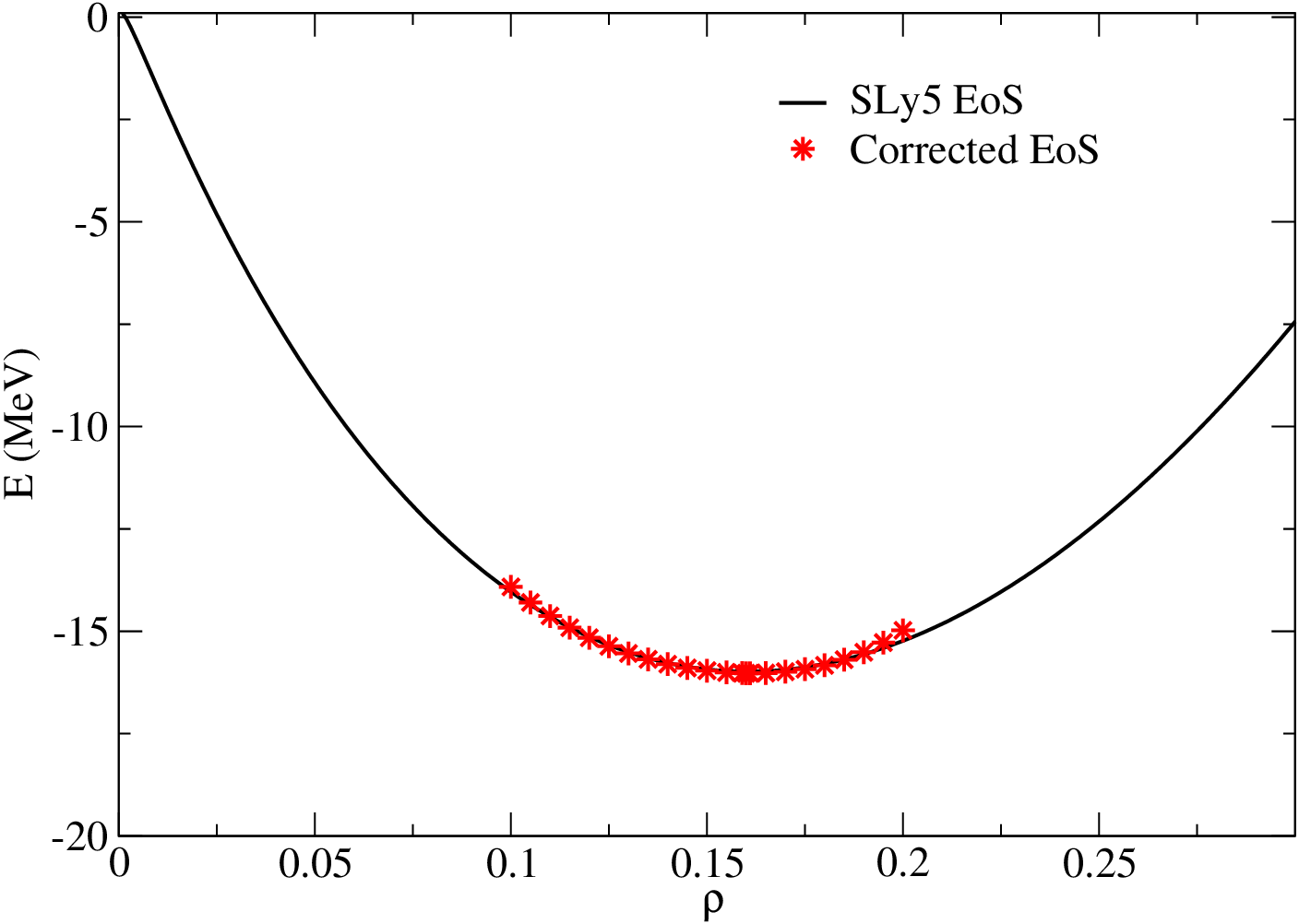}
\end{center}
\vspace{-.6cm}
\caption{\label{fig:fit3} (Color online) Fit4: Second-order refitted EoS compared with the SLy5-mean-field EoS for isospin symmetric nuclear matter.}
\end{figure}

\begin{figure}[h!]
\begin{center}
\includegraphics[scale=0.45 ]{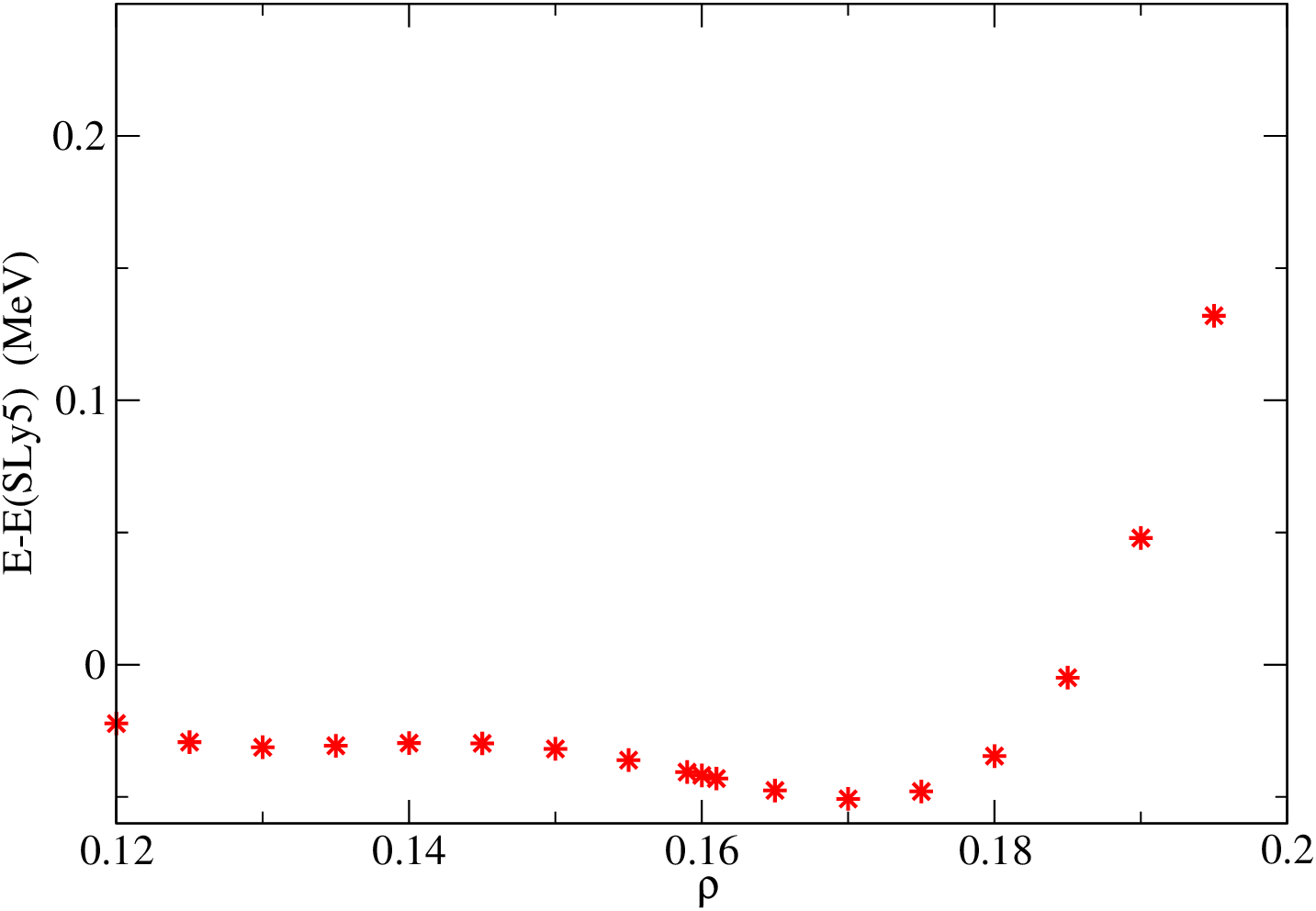}
\end{center}
\vspace{-.6cm}
\caption{\label{fig:eos-all1-diff} (Color online) Fit4: Difference between the refitted second-order and the mean-field EoS for isospin symmetric nuclear matter.}
\end{figure}

\begin{figure}[h!]
\begin{center}
\includegraphics[scale=0.45 ]{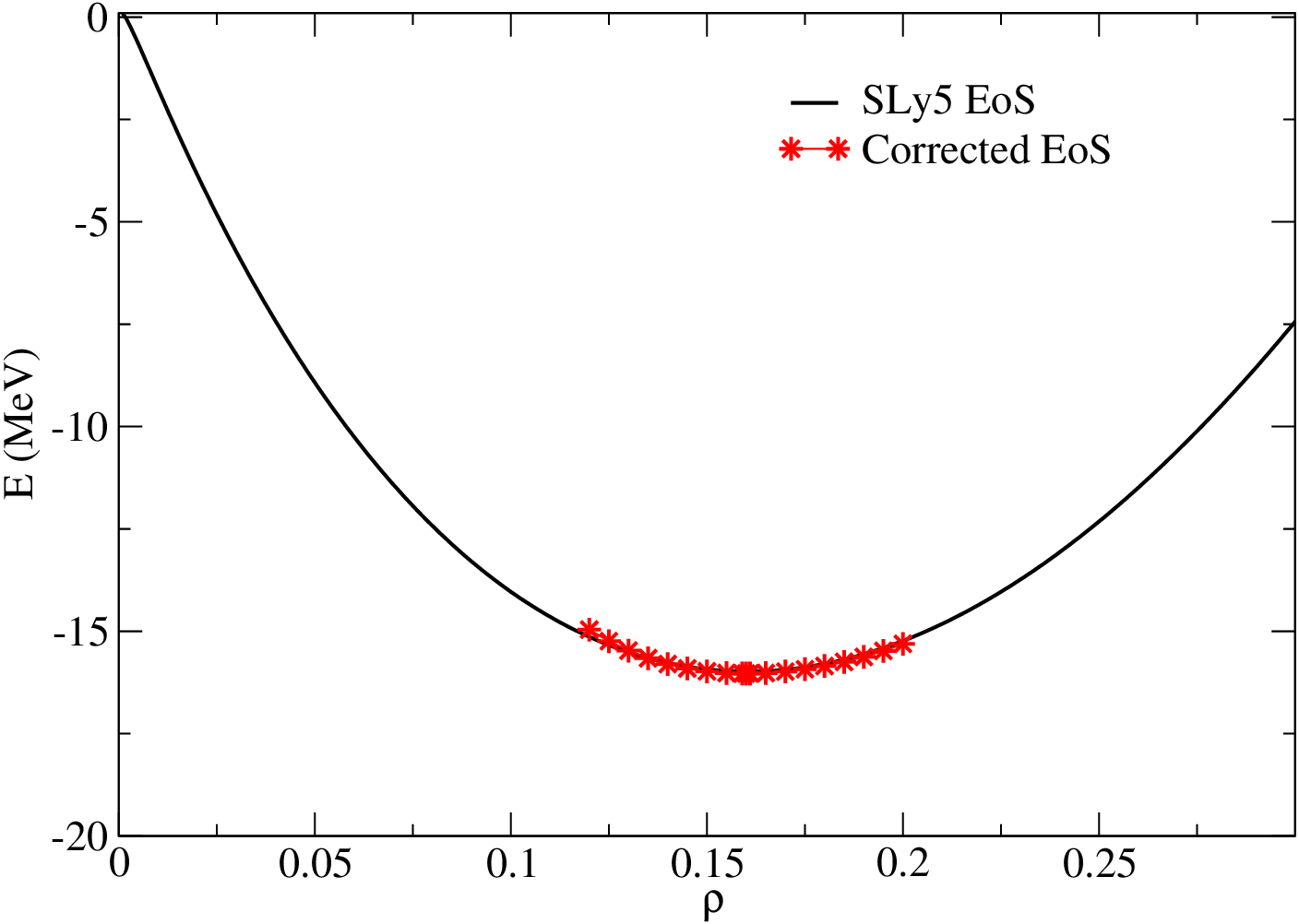}
\end{center}
\vspace{-.6cm}
\caption{\label{fig:fit4symm} (Color online) Fit5: Second-order refitted EoS compared with the SLy5-mean-field EoS for isospin symmetric nuclear matter.}
\end{figure}

\begin{figure}[h!]
\begin{center}
\includegraphics[scale=0.45 ]{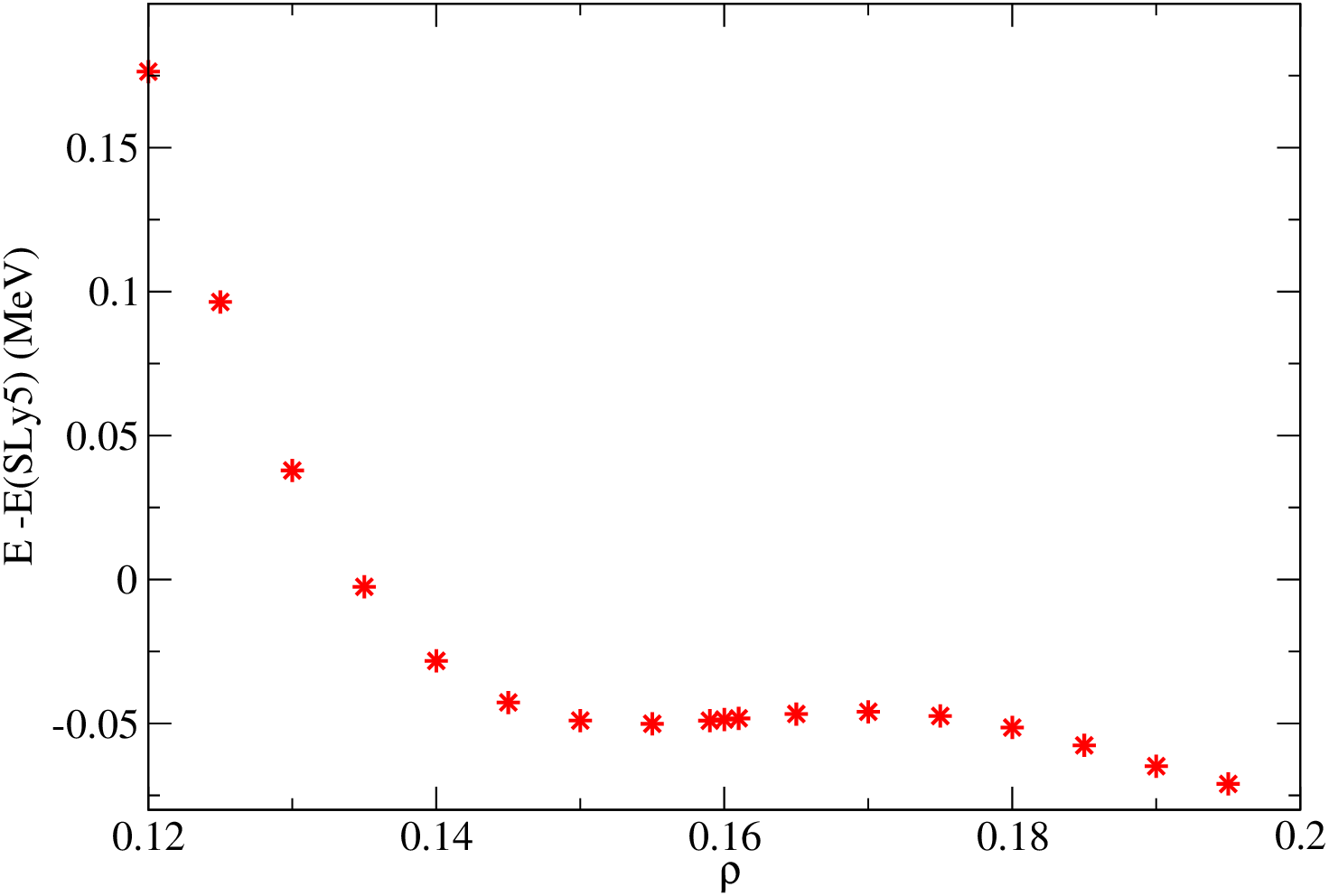}
\end{center}
\vspace{-.6cm}
\caption{\label{fig:eos-all3-diff-symm} (Color online) Fit5: Difference between the refitted second-order and the mean-field EoS for isospin symmetric nuclear matter.}
\end{figure}

\begin{figure}[h!]
\begin{center}
\includegraphics[scale=0.45]{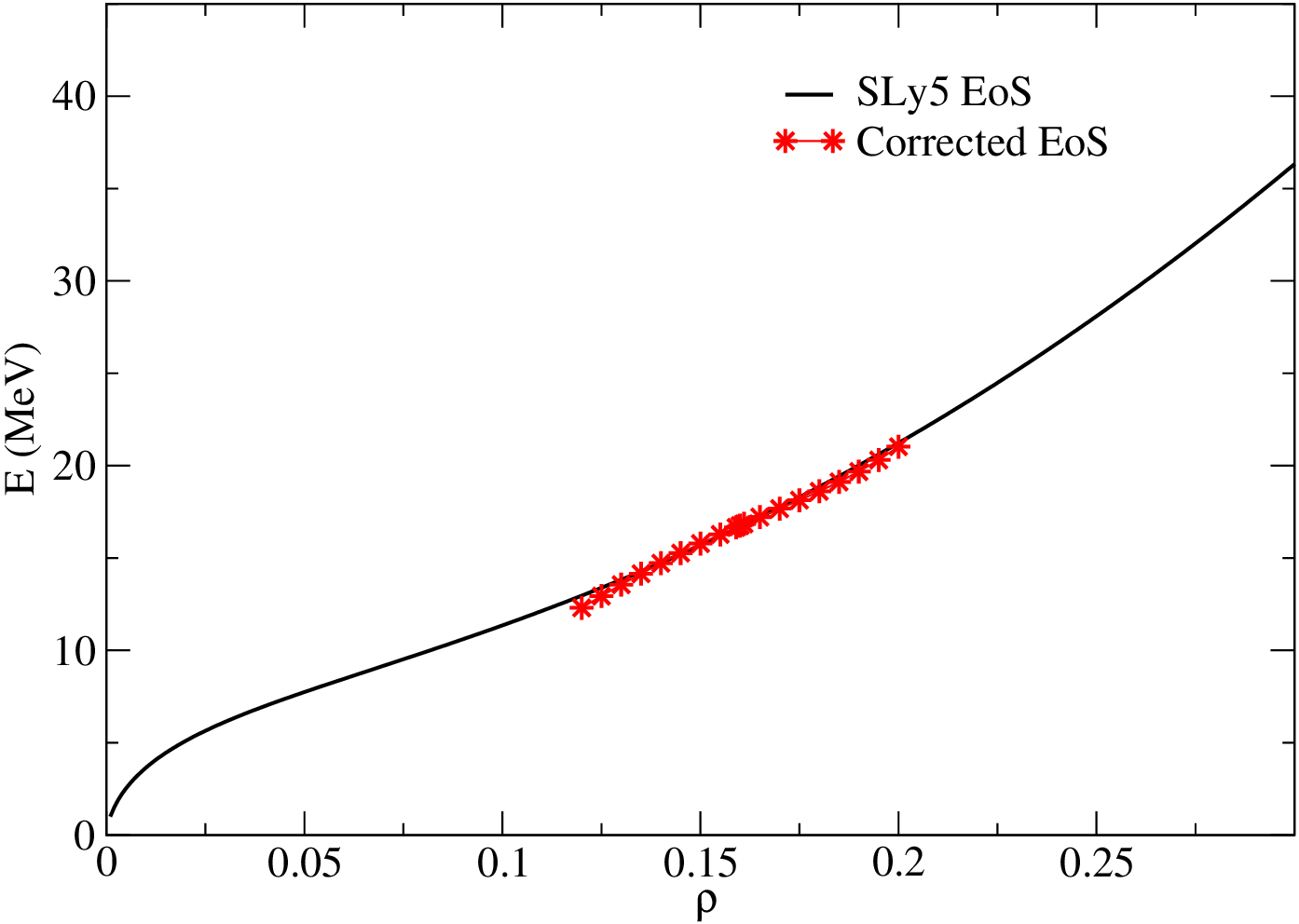}
\end{center}
\vspace{-.6cm}
\caption{\label{fig:fit4pure} (Color online) Fit5: Second-order refitted EoS compared with the SLy5-mean-field EoS for pure neutron matter.}
\end{figure}

\begin{figure}[h!]
\begin{center}
\includegraphics[scale=0.45]{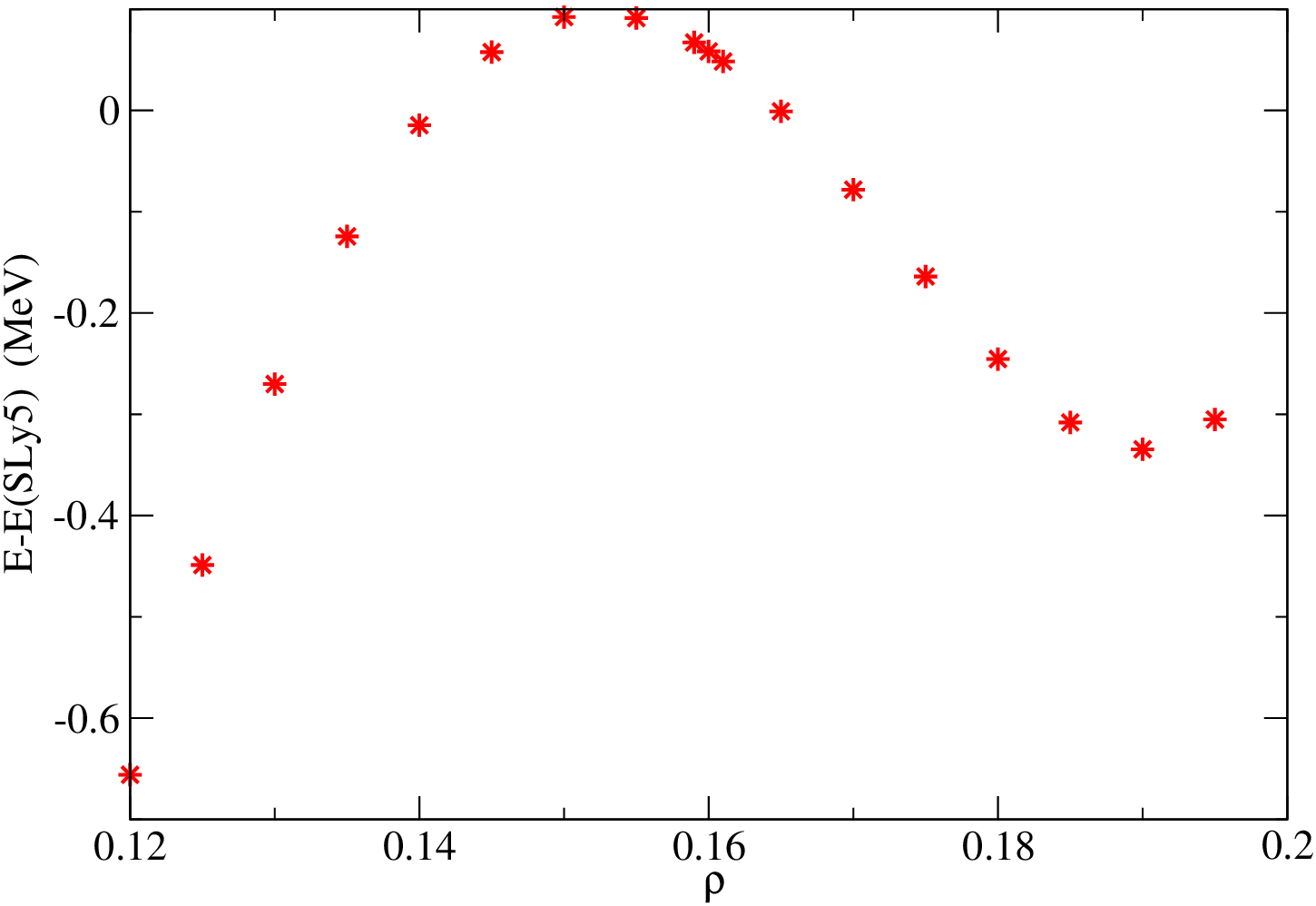}
\end{center}
\vspace{-.6cm}
\caption{\label{fig:eos-all3-diff-pure} (Color online) Fit5: Difference between the refitted second-order and the mean-field EoS for pure neutron matter.}
\end{figure}

In this section, we perform adjustments of the $18$ parameters of the Skyrme interaction in order to improve the nuclear bulk quantities of the EoS after the inclusion of second-order contributions from the N$^2$LO Skyrme interaction. We choose 18 equidistant reference points for densities ranging from 0.12 fm${}^{-3}$ to 0.200 fm${}^{-3}$ and two additional points located near the saturation density: 0.159 and 0.161 fm${}^{-3}$. The nuclear bulk quantities of the EoS provided by the phenomenological SLy5-interaction \cite{sly5} are taken as a benchmark \footnote{Of course, one could choose any benchmark other than the SLY5 mean-field EoS, for example, the Quantum Monte Carlo Calculations (QMC) of isospin symmetric nuclear matter \cite{QMC}.}. We use the $\chi^2$ minimization with the following definition:
\begin{equation}
\chi^2(Q_i)=\frac{1}{N}\sum_{i=1}^{N} \frac{(Q_i-Q_{i,ref})^2}{\Delta Q_i^2},
\end{equation}
where $N$ is the number of fitted points and the errors $\Delta Q_i$ are chosen equal to $1\%$ of the reference quantities $Q_{i,ref}$. \\
First of all, we adjust the parameters of the N$^2$LO Skyrme interaction in order to have a reasonable EoS of pure neutron matter. The refitted parameters are listed in Table \ref{tab:fits} (Fit1). In Fig.~\,\ref{fig:purealone}, we show the refitted EoS compared with the SLy5 mean-field EoS of pure neutron matter. In Fig.~\ref{fig:fit-pure-alone-diff}, the difference between the refitted second-order curve and the SLy5-mean-field curve for pure neutron matter is plotted as function of the density $\rho$, showing a maximum deviation of order $10^{-3}$. The quality of the fit is very good since the $\chi^2$ value is extremely small and equal to $0.000349$. \\
Next, we adjust the parameters of the N$^2$LO Skyrme interaction in order to have a reasonable EoS of isospin symmetric nuclear matter. Fig.~\ref{fig:symalone}, shows the refitted EoS compared with the SLy5 mean-field EoS of isospin symmetric nuclear matter and Fig.~\ref{fig:fit-symmetric-alone-diff} presents the difference between the refitted second-order curve and the SLy5-mean-field curve for isospin symmetric nuclear matter showing a deviation around the saturation density $\rho_0=0.160$ fm$^{-3}$ of order $10^{-2}$. The refitted parameters (Fit2) and the value of $\chi^2$ are listed in Table\,\ref{tab:fits}. The quality of the fit looks very good as indicated by the value of $\chi^2=0.0122$.  From table\,\ref{tab:bulk}, we note that the density and the energy per particle at saturation take reasonable values, however, the incompressibility modulus at saturation is equal to 196.036 MeV which is low compared to the SLy5 value: 229 MeV. Moreover, the isospin asymmetry energy at saturation is equal to 84.57 MeV which is far from 32 MeV and the effective mass at saturation takes negative value which is absolutely unreasonable. Although the quality of the fit is considered as remarkably good, the unrealistically negative value of effective nucleon mass and the large value of the isospin asymmetry energy display a serious problem. Therefore, the effective mass, the incompressibility, the isospin-asymmetry energy at saturation and the EoS of pure neutron matter should be constrained in the fitting procedure.\\
Next, we include in the adjustment of parameters the EoS of isospin symmetric nuclear matter and that of pure neutron matter. We use the same definition of $\chi^2$:
\begin{eqnarray} 
\chi^2=\frac{1}{2}\left[\chi^2_{(Q=E^{(2)}(k_F))}+\chi^2_{(Q=E_n^{(2)}(k_n))}\right].
\label{chi3}
\end{eqnarray}
In Fig.~\ref{fig:fit-symmetric-pure1}, we show the refitted EoS of isospin symmetric nuclear matter compared with the SLy5 mean-field EoS. In Fig.~\ref{fig:fit-symm-pure-diff1}, the difference between the refitted second-order curve and the SLy5-mean-field curve is plotted as function of the total density $\rho$ showing a small deviation of the energy per particle around the saturation density $\rho_0=0.160$ fm$^{-3}$ ($\sim 10^{-2}$). In Fig.~\ref{fig:fit-symm-pure2}, the refitted EoS of pure neutron matter is compared with the SLy5 mean-field EoS at different neutron densities $\rho$. In Fig.~\ref{fig:fit-symm-pure-diff2}, the deviation of the refitted second-order curve from the SLy5-mean-field curve is plotted as function of the density $\rho$ showing a deviation of order 10$^{-1}$. In Table\,\ref{tab:fits}, we list the values of the refitted parameters (Fit3) together with the  value of $\chi^2=0.353$ that show a good quality of the fit. From Table\,\ref{tab:bulk}, we note that the density, energy per particle and the incompressibility at saturation take reasonable values whereas the other bulk quantities: the effective mass and the isospin asymmetry energy take very unreasonable values.   \\
Next, we include in the adjustment of parameters the following nuclear bulk quantities: the saturation density $\rho_0$, the incompressibility modulus $K_{\infty}(\rho_0)$, the isospin asymmetry energy $A(\rho_0)$, the effective mass $m^*/m(\rho_0)$ and the EoS of isospin symmetric nuclear matter. In this case, the total $\chi^2$ is given by:
\begin{eqnarray} 
\chi^2=\frac{1}{5}\left[\chi^2_{(Q=E^{(2)}(k_F))}+\chi^2_{Q=\rho_0}+\chi^2_{Q=K_{\infty}(\rho_0)}+\chi^2_{Q=A(\rho_0)}+\chi^2_{Q=m^{*}/m(\rho_0)}\right],
\label{chi3}
\end{eqnarray}
where the five terms in eq.~(\ref{chi3}) represent the chi-squared deviations of the: EoS of isospin symmetric matter,  saturation density, incompressibility, isospin asymmetry energy, and nucleon effective mass respectively from their corresponding SLy5 values. In Fig.~\ref{fig:fit3}, we show the refitted EoS of isospin symmetric nuclear matter compared with the SLy5 mean-field EoS. The values of the refitted parameters (Fit4) and the value of $\chi^2$ are reported in Table\,\ref{tab:fits}. The difference between the refitted second-order curve and the SLy5-mean-field curve for isospin symmetric nuclear matter is plotted as function of the total density $\rho$ in Fig.~\ref{fig:eos-all1-diff}. The deviation of the energy per particle around the saturation density $\rho_0=0.160$ fm$^{-3}$ is of order $10^{-2}$ and the value of $\chi^2=0.271$ show a good quality of the fit. Moreover, we note from Table\,\ref{tab:bulk} that the other adjusted bulk quantities are very well reproduced as they take reasonable values. Therefore a very good reproduction of the realistic nuclear matter saturation curve is obtained for densities ranging from 0.12 to 0.2 fm$^{-3}$ with a very good quality of the fit. \\
Finally, we include on top of the previous fit the EoS of pure neutron matter. Now, the total $\chi^2$ is given by:
\be 
\chi^2=\frac{1}{6}\left[\chi^2_{Q=E^{(2)}(k_F)}+\chi^2_{Q=E_n^{(2)}(k_n)}+\chi^2_{Q=\rho_0}+\chi^2_{Q=K_{\infty}(\rho_0)}+\chi^2_{Q=A(\rho_0)}+\chi^2_{Q=m^{*}/m(\rho_0)}\right].
\ee
In Figs.~\ref{fig:fit4symm} and \ref{fig:fit4pure}, we show respectively the refitted EoS of isospin symmetric nuclear matter and pure neutron matter compared with their corresponding SLy5 mean-field EoS. The values of the refitted parameters (Fit5) and the value of $\chi^2$ are reported in Table\,\ref{tab:fits}. In Figs.~\ref{fig:eos-all3-diff-symm} and ~\ref{fig:eos-all3-diff-pure}, we show the difference between the refitted second-order curve and the SLy5-mean-field curve for both isospin symmetric and pure neutron matter. The deviation of the EoS of isospin symmetric matter around the saturation density $\rho_0=0.160$ fm$^{-3}$ is of order $10^{-2}$. The value of $\chi^2=1.102$ and the values of the other bulk quantities show that a very good reproduction of the realistic nuclear matter saturation curve is be obtained for densities ranging from 0.12 to 0.2 fm$^{-3}$.

\begin{table}[h!]
\caption{\label{tab:fits} Parameter sets with different kind of fits. The values of $\chi^2$ are shown in the last row.}
\begin{ruledtabular}
\begin{tabular}{lccccccc}
\multicolumn{1}{c}{Parameters}& \multicolumn{1}{c}{Fit1} & \multicolumn{1}{c}{Fit2}&\multicolumn{1}{c}{Fit3}& \multicolumn{1}{c}{Fit4} & \multicolumn{1}{c}{Fit5}  \\
\hline
$t_0$ [MeV\,fm$^3$]  &-838.922       & -936.148 & -1688.619& -1058.790  & -1629.806     \\
$x_0$                     &  0.0329 & 0.0254  &0.463& -0.0747     & 0.216    \\                     
$t_1$ [MeV\,fm$^5$]   &    46.488  &  228.963 & 1961.006 &2119.184  & 2579.990      \\
$x_1$         & 3.729& 1.275  & 1.099& -0.0921  & 0.244    \\
$t_2$ [MeV\,fm$^5$] & -1585.881 & -2755.314  &-2098.979&   -822.011    & -2311.918      \\
$x_2$ & -0.506 & 0.161  & -0.359& 1.229     &  -0.816     \\ 
$W_0$ [MeV\,fm$^5$] & -1.535&  -17.841    & 2.047& 100.000      &  53.524    \\ 
\hline
$t_6$ [MeV\,fm$^9$]    & 10.252 & 29.828 &-125.735&  -76.851    &  -140.262    \\
$x_6$           & -0.0846&  -1.176  &1.136&  -0.408      &  0.365      \\
$t_7$ [MeV\,fm$^9$]    & -57.991 & -314.272   &-496.945&   -761.790   & -458.858       \\  
$x_7$   & 1.232& 0.00286  &1.158&  -0.0734    &  0.134     \\
$t_8$ [MeV\,fm$^9$]  &  185.442&  435.504  &406.191& 224.694     &  570.729       \\
$x_8$ &  -0.716& 0.321    &-0.649&   1.165     &  -0.930   \\
$t_9$ [MeV\,fm$^9$]     &  145.935& 268.903     & 36.618& -468.737    & -180.124      \\
$x_9$     &  0.288 &  0.101 &-3.761& 0.108     &  0.530      \\
$t_{10}$   [MeV\,fm$^9$]           &  -0.293 &  4.103   &0.00170&  -23.258        & -8.311    \\
$t_{11}$  [MeV\,fm$^9$]   &  -0.00260&  -0.002600    &0.00255&   -5.040    &   -1.964         \\
$t_{12}$  [MeV\,fm$^9$] &  26.704 &  -4.325   &-0.592&  -18.175    & 21.891       \\
$\chi^{2}$    & 0.000349  & 0.0122  &0.353&  0.271      &  1.102 
\end{tabular}
\end{ruledtabular}
\end{table}

\begin{table}[h!]
\caption{\label{tab:bulk} Values of the second-order nuclear bulk quantities at saturation with different fits compared with the SLy5 mean-field bulk quantities.}
\begin{ruledtabular}
\begin{tabular}{lcccccccc}
\multicolumn{1}{c}{$Q(\rho_0$)}&\multicolumn{1}{c}{SLy5} & \multicolumn{1}{c}{Fit1} & \multicolumn{1}{c}{Fit2}& \multicolumn{1}{c}{Fit3}& \multicolumn{1}{c}{Fit4} & \multicolumn{1}{c}{Fit5}  \\
\hline
$\rho_0$ [fm$^{-3}]$        &   0.161  &-& 0.159 &0.158& 0.160 & 0.160       \\ 
$E_0$ [MeV]        &-15.983 & -&-15.96 & -16.009&-16.025 & -16.032      \\                
$K_{\infty}$ [MeV\,fm$^5$] &  229.203     & -& 196.036 &227.053 &228.784  & 232.960            \\
$A_0 [MeV]$                 & 31.992              &-&  84.570&70.035 & 31.993  &  32.210     \\
$ m^*/m$         &0.697 &-& -0.105 &-0.41&0.697  &   0.705        
\end{tabular}
\end{ruledtabular}
\end{table}
\vspace{-0.5cm}

\section{Conclusions and outlook}
In this work, we have considered the N$^2$LO Skyrme interaction that is an extension to the standard Skyrme interaction by including all types of fourth-derivative contact potentials without constrainting the interaction to be gauge-invariant since the latter is not explicitly required from basic principles. The N$^2$LO Skyrme interaction includes central, non-central and higher-order spin-orbit (non-gauge invariant) contact forces. We have derived analytically  the first-order corrections to the nuclear bulk quantities: EoS of isospin symmetric and pure neutron matter, density-dependent effective mass, isospin-asymmetry energy, pressure and incompressibility modulus. Moreover, we have considered the second-order diagrams or the (2p-2h) excitations beyond the mean-field level and derived analytically the expressions for the second-order contributions of the various nuclear bulk quantities using the N$^2$LO Skyrme interaction. Due to the zero-range character of the interaction, UV divergences are present in the second-order contributions. In this case, we have used the momentum cutoff (MC) scheme and the techniques of dimensional regularization with the minimal subtraction scheme (DR/MS) to renormalize all divergent integrals. We have seen, in the MC scheme, that some of the divergent terms cannot be renormalized by a redefinition of the existing parameters and hence a problem of perturbativity of the Skyrme interaction arises. On the other hand, DR/MS automatically sets all power divergent integrals to zero and hence all the second-order corrections are renormalized. Moreover, we have found that although the higher-order spin-orbit potential do not contribute at first-order, they do contribute at second order to the various nuclear bulk quantities and therefore such terms are important in the calculation of nuclear many-body quantities.\\
The 18 parameters of the N$^2$LO Skyrme interactions are adjusted in such a way the EoS curve is well reproduced. We have performed a global fit that readjusts the various nuclear bulk quantities to specific benchmark chosen to be the SLy5 mean-field EoS curve. We have shown that, besides the too strong density dependence of some second-order contributions, one can obtain a very good reproduction of a realistic nuclear matter saturation curve with all the nuclear bulk quantities in the density region $0.120<\rho<0.20\;$ fm$^{-3}$. Therefore, from these findings one can conclude that the going beyond the standard Skyrme interaction by including the fourth-derivative contact terms, we are able to reproduce the realistic nuclear matter saturation curve. Work is in progress to include the contributions of the tensor forces, the density-dependent with the rearrangement terms and the calculations of the many-body Landau parameters \cite{moghrabi} at N$^2$LO.

\section*{Acknowledgements}
This work was supported by the doctoral school of sciences and technology in the Lebanese university. I thank  H. Zaraket, N. Kaiser and A. Pastore for informative and fruitful discussions.

\newpage

\appendix*
\section{}
In this appendix, we display the analytical expressions of the second-order corrections to the various nuclear bulk quantities. We start by showing the expressions for that to the EoS of isospin symmetric nuclear matter.
\be e_{1s}&=&\frac{1}{324324000}\bigg\{1152t_6^2\left(1+x_6^2\right)\left(129593-8895 \ln 2\right)+36t_7^2\left(1+x_7^2\right)\left(19697-2670 \ln 2\right)+[3t_9^2\left(1+x_9^2\right)+3t_{12}^2\\&&\:-2t_9t_{12}\left(1-2x_9\right)]\left(948097-125760 \ln 2\right)
+8\left[6t_6t_7\left(1+x_6x_7\right)+t_8^2\left(5+5x_8^2+8x_8\right)\right]\left(265541-29940 \ln 2\right)\\
&&\:+200\left[3t_6t_9\left(1+x_6x_9\right)-t_6t_{12}\left(1-2x_6\right)+6t_{10}^2\right]\left(58979-6144 \ln 2\right)
+2\left[3t_7t_9\left(1+x_7x_9\right)-t_7t_{12}\left(1-2x_7\right)+2t_{11}^2\right]\\&&\:\times\left(198193-27840 \ln 2\right)\bigg\},\\
e_{2s}&=&\frac{1}{3603600}\bigg\{12t_1t_6\left(1+x_1x_6\right)\left(81799-7320 \ln 2\right)+[51t_1t_7\left(1+x_1x_7\right)+34t_2t_8\left(5+5x_2x_8+4x_8+4x_2\right)
\\
&&\:+132t_1t_9\left(1+x_1x_9\right)-44t_1t_{12}\left(1-2x_1\right)+1056W_0t_{10}]\left(919-120 \ln 2\right)\bigg\},\\
e_{3s}&=&\frac{1}{166320}\bigg\{12t_0t_6\left(1+x_0x_6\right)\left(4943-564 \ln 2\right)+3t_0t_7\left(1+x_0x_7\right)\left(1033-156 \ln 2\right)+4\left[3t_0t_9\left(1+x_0x_9\right)-t_0t_{12}\left(1-2x_0\right)\right]\\&&\:\times\left(631-102 \ln 2\right)\bigg\}.
\ee
Similarly, the second-order corrections to the EoS of pure neutron matter can be written in terms of $e_{in}$:
\be 
e_{1n}&=&\frac{1}{324324000}\bigg\{384t_6^2\left(1-x_6\right)^2\left(129593-8895 \ln 2\right)+12t_7^2\left(1-x_7\right)^2\left(19697-2670 \ln 2\right)\\&&\:
+\left(t_9-t_9x_9-t_{12}\right)^2\left(948097-125760 \ln 2\right)+
8\left[2t_6t_7\left(1-x_6\right)\left(1-x_7\right)+3t_8^2\left(1+x_8\right)^2\right]\left(265541-29940 \ln 2\right)\\&&\:
+200\left[t_6\left(1-x_6\right)\left(t_9-t_9x_9-t_{12}\right)+4t_{10}^2\right]\left(58979-6144 \ln 2\right)\\
&&\:+2\left[t_7\left(1-x_7\right)\left(t_9-t_9x_9-t_{12}\right)\right]\left(198193-27840 \ln 2\right)\bigg\},\\
e_{2n}&=&\frac{1}{3603600 }\bigg\{4t_1t_6\left(1-x_1\right)\left(1-x_6\right)\left[81799-7320 \ln 2\right)+
17\left[t_1t_7\left(1-x_1\right)\left(1-x_7\right)+6t_2t_8\left(1+x_2\right)\left(1+x_8\right)\right]\left(919-120 \ln 2\right)\\
&+&44\left[t_1\left(1-x_1\right)\left(t_9-t_9x_9-t_{12}\right)+16W_0t_{10}\right]\left(919-120 \ln 2\right)\bigg\},\\
e_{3n}&=&\frac{1}{166320}\bigg\{4t_0t_6\left(1-x_0\right)\left(1-x_6\right)\left(4943-564 \ln 2\right)+t_0t_7\left(1-x_0\right)\left(1-x_7\right)\left(1033-156 \ln 2\right)
\\
&+&4t_0\left(1-x_0\right)\left(t_9-t_9x_9-t_{12}\right)\left(631-102 \ln 2\right)\bigg\}.
\ee
The second-order contribution to the isospin asymmetry energy $A(k_F)$ is written in terms of $a_i$ where their expressions read:
\newpage
\be 
a_1&=&\frac{3}{8}\bigg\{t_6^2\left(1-x_6\right)^2\left(\frac{8293952}{2027025}-\frac{37952 \ln 2}{135135}\right)+2t_6^2\left(1+x_6+x_6^2\right)\left(-\frac{98008}{6081075}-\frac{12352 \ln 2}{405405}\right)\\
&+&t_7^2\left(1-x_7\right)^2\left(\frac{39394}{2027025}-\frac{356 \ln 2}{135135}\right)+2t_7^2\left(1+x_7+x_7^2\right)\left(\frac{629}{162162}-\frac{356 \ln 2}{135135}\right) \\
&+&\left(t_9-t_9x_9-t_{12}\right)^2\left(\frac{948097}{12162150}-\frac{4192 \ln 2}{405405}\right)+2\left[t_9^2\left(1+x_9+x_9^2\right)+t_{12}^2+t_9x_9t_{12}\right]\left(-\frac{11471}{1351350}+\frac{128 \ln 2}{135135}\right)\\
&+&2t_{11}^2\left(-\frac{446}{552825}-\frac{8 \ln 2}{36855}\right)+
\left[2t_6t_7\left(1-x_6\right)\left(1-x_7\right)+3t_8^2\left(1+x_8\right)^2\right]\left(\frac{1062164}{6081075}-\frac{7984 \ln 2}{405405}\right)\\&+&2\left[t_6t_7\left(2+2x_6x_7+x_6+x_7\right)+t_8^2\left(1+x_8+x_8^2\right)\right]\left(\frac{123148}{6081075}-\frac{4448 \ln 2}{405405}\right)\\
&+&2t_6\left(1-x_6\right)\left(t_9-t_9x_9-t_{12}\right)\left(\frac{117958}{243243}-\frac{4096 \ln 2}{81081}\right) +2t_7\left(1-x_7\right)\left(t_9-t_9x_9-t_{12}\right)\left(\frac{198193}{12162150}-\frac{928 \ln 2}{405405}\right)\\&+&2\left[t_6t_9\left(2+2x_6x_9+x_6+x_9\right)+t_6x_6t_{12}\right]\left(-\frac{45832}{1216215}+\frac{272 \ln 2}{81081}\right)\\&+&2\left[t_7t_9\left(2+2x_7x_9+x_7+x_9\right)+t_7x_7t_{12}\right]\left(-\frac{223}{552825}-\frac{4 \ln 2}{36855}\right)+\frac{16}{12285}t_{10}^2\left(2863-300 \ln 2\right)\\
&+&2t_6t_8\left(2+2x_6x_8+x_8+x_6\right)\left(\frac{976 \ln 2}{31185}-\frac{81626}{467775}\right)+2t_7t_8\left(2+2x_7x_8+x_8+x_7\right)\left(\frac{68 \ln 2}{10395}-\frac{5921}{311850}\right)\\
&+&2\left[t_8t_9\left(2+2x_8x_9+x_9+x_8\right)+4t_{10}t_{11}\right]\left(\frac{8 \ln 2}{6237}-\frac{919}{93555}\right)\bigg\},
\ee

\be 
a_2&=&\frac{3}{4}\bigg\{\frac{t_1}{2}t_6\left(1-x_1\right)\left(1-x_6\right)\left(\frac{163598}{93555}-\frac{976 \ln 2}{6237}\right)+\frac{t_1}{2}t_6\left(2+2x_1x_6+x_6+x_1\right)\left(-\frac{2561}{31185}-\frac{248 \ln 2}{10395}\right) \\
&+&\frac{t_1}{2}t_7\left(1-x_1\right)\left(1-x_7\right)\left(\frac{15623}{187110}-\frac{68 \ln 2}{6237}\right) +3t_2t_8\left(1+x_2\right)\left(1+x_8\right)\left(\frac{15623}{187110}-\frac{68 \ln 2}{6237}\right) \\&+&\left[\frac{t_1}{2}t_7\left(2+2x_1x_7+x_7+x_1\right)+t_2t_8\left(2+2x_2x_8+x_2+x_8\right)\right]\left(\frac{1867}{224532}-\frac{734 \ln 2}{93555}\right)
\\
&+&\frac{t_1}{2}\left(1-x_1\right)\left(t_9-t_9x_9-t_{12}\right)\left(\frac{1838}{8505}-\frac{16 \ln 2}{567}\right)\\&+&\left[\frac{t_1}{2}t_9\left(2+2x_1x_9+x_1+x_9\right)+\frac{t_1x_1}{2}t_{12}\right]\left(-\frac{691}{25515}+\frac{16 \ln 2}{8505}\right)+\frac{4}{25515}W_0t_{10}\left(10337-1392 \ln 2\right)\\
&+&t_0t_8\left(2+2x_0x_8+x_8+x_0\right)\left(\frac{16 \ln 2}{945}-\frac{124}{2835}\right)\\
&+&\left[\frac{t_1}{2}t_8\left(2+2x_1x_8+x_8+x_1\right)+t_2t_6\left(2+2x_2x_6+x_6+x_2\right)\right]\left(\frac{188 \ln 2}{8505}-\frac{4279}{51030}\right)\\
&+&t_2t_7\left(2+2x_2x_7+x_7+x_2\right)\left(\frac{13 \ln 2}{2835}-\frac{701}{68040}\right)\\
&+&\left[t_2t_9\left(2+2x_2x_9+x_9+x_2\right)+t_2x_2t_{12}+W_0t_{12}\right]\left(\frac{8 \ln 2}{8505}-\frac{229}{51030}\right)\bigg\},
\ee

\be 
a_3&=&\frac{3}{4}\bigg\{t_0t_6\left(1-x_0\right)\left(1-x_6\right)\left(\frac{19772}{25515}-\frac{752 \ln 2}{8505}\right)+t_0t_6\left(2+2x_0x_6+x_6+x_0\right)\left(-\frac{416}{5103}-\frac{176 \ln 2}{8505}\right)\\
&+&t_0t_7\left(1-x_0\right)\left(1-x_7\right)\left(\frac{1033}{25515}-\frac{52 \ln 2}{8505}\right)+t_0t_7\left(2+2x_0x_7+x_7+x_0\right)\left(\frac{88}{25515}-\frac{52 \ln 2}{8505}\right)
\\
&+&t_0\left(1-x_0\right)\left(t_9-t_9x_9-t_{12}\right)\left(\frac{2524}{25515}-\frac{136 \ln 2}{8505}\right)+\left[t_0t_9\left(2+2x_0x_9+x_9+x_0\right)+t_0x_0t_{12}\right]\left(-\frac{482}{25515}+\frac{8 \ln 2}{8505}\right)\bigg\}.
\ee
The second-order corrections to the nucleonic effective mass can be written in terms of $r_i$:
\be 
r_1&=&\frac{1}{10810800}\bigg\{192t_6^2\left(1+x_6^2\right)\left(368957-134430 \ln 2\right)+792t_7^2\left(1+x_7^2\right)\left(649-615 \ln 2\right)\\
&+&3\left[3t_9^2\left(1+x_9^2\right)+3t_{12}^2-2t_9t_{12}\left(1-2x_9\right)\right]\left(144751-79680 \ln 2\right)
+16 t_{11}^2\left(25393-17040 \ln 2\right)\\
&+&136\left[6t_6t_7\left(1+x_6x_7\right)+t_8^2\left(5+5x_8^2+8x_8\right)\right]\left(9667-6780 \ln 2\right)\\
&+&160\left[3t_6t_9\left(1+x_6x_9\right)-t_6t_{12}\left(1-2x_6\right)\right]\left(33119-15360 \ln 2\right)\\
&+&8\left[3t_7t_9\left(1+x_7x_9\right)-t_7t_{12}\left(1-2x_7\right)\right]\left(25393-17040 \ln 2\right)+960t_{10}^2\left(33119-15360 \ln 2\right)\bigg\},\\
r_2&=&\frac{1}{166320}\bigg\{12t_1t_6\left(1+x_1x_6\right)\left(35173-20712 \ln 2\right)
+3\left[3t_1t_7\left(1+x_1x_7\right)+2t_2t_8\left(5+5x_2x_8+4x_8+4x_2\right)\right]\left(3159-3032 \ln 2\right)\\
&&\:+968\left[3t_1t_9\left(1+x_1x_9\right)-t_1t_{12}\left(1-2x_1\right)+24 W_0t_{10}\right]\left(17-12 \ln 2\right),\\
r_3&=&\frac{1}{11340}\bigg\{24t_0t_6\left(1+x_0x_6\right)\left(911-942 \ln 2\right)+48t_0t_7\left(1+x_0x_7\right)\left(313-426 \ln 2 \right)
+\left[3t_0t_9\left(1+x_0x_9\right)-t_0t_{12}\left(1-2x_0\right)\right]\\&&\:\times
\left(881-1032 \ln 2\right)\bigg\}.
\ee
The second-order corrections to the various nuclear bulk quantities from the standard Skyrme interaction have been evaluated in \cite{kaiser1}. We start with the EoS of isospin symmetric nuclear and pure neutron matter:
\be 
\Delta E^{(2)}_{\text{Sk2}}(k_F) &=&{M k_F^4 \over 280 \pi^4} \bigg\{ 
3t_0^2(1+x_0^2)(11-2\ln 2) +{2k_F^2 \over 9}t_0 t_1(1+x_0 x_1)
(167-24\ln 2) +{k_F^4 \over 396} t_1^2(1+x_1^2)(4943-564\ln 2)\nonumber\\&&
+{k_F^4 \over 1188}t_2^2(5+8x_2+5 x_2^2)(1033-156\ln 2)+{4k_F^4 \over 99}W_0^2 (631-102\ln 2) \bigg\}\\
\Delta E^{(2)}_{\text{Sk2}}(k_n) &=&{M k_n^4 \over 280\pi^4}\bigg\{ 
t_0^2(1-x_0)^2(11-2\ln 2) +{2k_n^2 \over 27}t_0 t_1(1-x_0)(1-x_1)
(167-24\ln 2) +{k_n^4 \over 1188} t_1^2(1-x_1)^2(4943-564\ln 2)
\nonumber \\ &&+{k_n^4 \over 396}t_2^2(1+x_2)^2(1033-156\ln 2)
+{8k_n^4 \over 297}W_0^2 (631-102\ln 2) \bigg\}\,.
\ee
The second-order contributions to the isospin asymmetry energy can be written as: 
\be 
\Delta A^{(2)}_{\text{Sk2}}(k_F)&=&{M k_F^4 \over 60\pi^4}\bigg\{ 
t_0^2\big[(1+x_0^2)(1-2\ln 2) -10x_0\big]  + {k_F^2 \over 42}
t_0 t_1\big[4(1+x_0x_1)(51-22\ln 2) +7(x_0+x_1)(4\ln 2-57)\big]\nonumber \\ && 
+{k_F^4 \over 63}t_1^2\Big[(1+x_1^2)\Big({1301 \over 6}-46\ln 2\Big) 
+x_1(48\ln 2-607)\Big] +{k_F^4 \over 567}t_2^2\Big[
5(1+x_2^2)\Big({655 \over 2}-78\ln 2\Big) +x_2(3187-624\ln 2)\Big]\nonumber 
\\ && + {k_F^2 \over 6} t_0 t_2(2+x_0+x_2+2x_0x_2)(4\ln 2-7) + 
{2k_F^4 \over 63}t_1 t_2(2+x_1+x_2+2x_1x_2)(12\ln 2-31)\nonumber \\ && 
+{4k_F^4 \over 189}W_0^2 (761-132\ln 2)\bigg\}\,.
\ee
Finally, the second-order correction to the nucleonic effective mass reads:
\be 
\Delta R^{(2)}_{\text{Sk2}}(k_F)&=& {M^2 k_F^2 \over 10\pi^4} \bigg\{t_0^2(1
+x_0^2)(1-7\ln 2)+{k_F^2\over 14}t_0t_1(1+x_0 x_1)(47-104\ln 2) 
+{k_F^4\over 378}t_1^2(1+x_1^2)(911-942\ln 2)\nonumber \\ &&+{k_F^4\over 1134}t_2^2(5+8x_2
+5x_2^2)(313-426\ln 2) +{k_F^4\over 189}W_0^2(881-1032\ln 2)
\bigg\} .
\ee
\end{document}